\documentclass[12pt,preprint]{aastex}
\shorttitle{UV Properties of SDSS Quasars}
\shortauthors{Trammell et al.}

\received{2006 July 7}
\begin{document}

\slugcomment{1 November 2006 draft}
\title{The UV Properties of SDSS Selected Quasars} 

\author{George B. Trammell \altaffilmark{1,2,3}, 
Daniel E. Vanden Berk \altaffilmark{1}, 
Donald P. Schneider \altaffilmark{1},
Gordon T. Richards \altaffilmark{4,5},
Patrick B. Hall \altaffilmark{6},
Scott F. Anderson \altaffilmark{7},
\& J. Brinkmann \altaffilmark{8}} 

\altaffiltext{1}{Department of Astronomy and Astrophysics, The
Pennsylvania State University, 525 Davey Laboratory, University Park,
PA 16802}

\altaffiltext{2}{Department of Astronomy, Box 400325, University of
Virginia, Charlottesville, VA 22904}

\altaffiltext{3}{e-mail: trammell@astro.psu.edu}

\altaffiltext{4}{Department of Physics and Astronomy, The Johns
Hopkins University, 3400 North Charles Street, Baltimore, MD
21218-2686}

\altaffiltext{5}{Department of Physics, Drexel University, 3141 Chestnut
Street, Philadelphia, PA 19104}

\altaffiltext{6}{Department of Physics and Astronomy, York University,
4700 Keele Street, Toronto, ON, M3J 1P3, Canada}

\altaffiltext{7}{Astronomy Department, Box 351580, University of Washington, 
Seattle, WA 98195}

\altaffiltext{8}{Apache Point Observatory, Box 59, Sunspot, NM 88349}

\begin{abstract}

We present an analysis of the broadband UV and optical properties of
$z \lesssim 3.4$ quasars matched in the \textit{Galaxy Evolution
Explorer} ({\it GALEX}) General Data Release 1 (GR1) and the
Sloan Digital Sky Survey Data Release 3 (SDSS DR3).  Of the
6371 SDSS DR3 quasars covered by 204 {\it GALEX} GR1 tiles, 5380
(84\%) have near-UV detections, while 3034 (48\%) have both near-UV
and far-UV detections using a matching radius of 7\arcsec.  Most of
the DR3 sample quasars are detected in the near-UV until $z \sim 1.7$,
with the near-UV detection fraction dropping to $\sim 50\%$ by $z \sim
2$.  Statistical tests performed on the distributions of
non-detections indicate that the optically selected quasars missed in
the UV tend to be optically faint or at high redshift.  The {\it
GALEX} positions are shown to be consistent with the SDSS astrometry
to within an rms scatter of 0.6-0.7{\arcsec} in each coordinate, and
the empirically determined photometric errors from multi-epoch {\it
GALEX} observations significantly exceed the Poissonian errors quoted
in the GR1 object catalogs.  The UV-detected quasars are well
separated from stars in UV/optical color-color space, with the
UV/optical relative colors suggesting a marginally detected population
of reddened objects due to absorption along the line of sight or due
to dust associated with the quasar.  The resulting spectral energy
distributions (SEDs) cover $\sim350-9000$ {\AA} (rest frame), where
the overall median SED peaks near the Ly $\alpha$ emission line, as
found in other UV quasar studies.  The large sample size allows us to
construct median SEDs in small bins of redshift and luminosity, and we
find the median SED becomes harder (bluer) at UV wavelengths for
quasars with lower continuum luminosity.  The detected UV/optical flux
as a function of redshift is qualitatively consistent with attenuation
by intervening Lyman-absorbing clouds.  A table containing the 6371
DR3 quasars covered by {\it GALEX} GR1 and the results of the matching
is available in the electronic edition of the journal, and a table
containing the overall median SED is also available.

\end{abstract}


\section{INTRODUCTION} 
\label{intro}

One of the primary goals of large-scale UV and optical sky surveys is
to create large sample sizes, crucial for a statistically significant
measurement and interpretation of the ubiquitous properties displayed
by active galactic nuclei (AGN), such as broad emission lines and
power-law continua.  In quasars, a large fraction of the bolometric
luminosity is emitted in a strong, broad feature that begins to
dominate the spectral energy distribution (SED) at the bluest optical
wavelengths \citep[the `big blue bump'; hereafter BBB;][]{Sanders
1989} and appears to extend shortward to the current limits of UV
satellite data ($\sim 100$ {\AA}).  Previous work that interprets the
rise in the continuum as the thermal signature of a large and hot
accretion disk surrounding a central supermassive black hole
\citep[e.g.,][]{Shields 1978, MS 1982} has argued that the BBB should
peak in the extreme ultraviolet (EUV), shortward of the Lyman limit at
912 {\AA} \citep{MF 1987, Hubeny 2000, Frank 2002}.  Superimposed on
the BBB is yet another feature (the `small blue bump'; hereafter SBB)
attributed to Balmer continuum and complex Fe II emission from the
broad line region \citep{Grandi 1981, Wills 1985}.  The BBB and SBB
are characteristic features of quasar emission; therefore UV and
optical observations are a critical probe of the physics of the inner
regions of quasars.  Deep UV observations have the added capability of
sampling the ionizing flux, which gives rise to the powerful emission
lines that are a trademark of quasar spectra.

\citet{Zheng 1997} have used \textit{Hubble Space Telescope}
(\textit{HST}) observations to construct composite rest-frame UV
spectra from 101 AGN with $z > 0.33$ that display a break in the
UV/optical power-law continuum near $\sim 1000$ {\AA} (specifically,
the EUV continuum blueward of the break is softer than the continuum
redward of the break).  This result was confirmed by \citet{Telfer
2002} for a larger sample of 184 AGN with redshifts in the range $0.33
< z < 3.6$, who also demonstrated no evolution of the EUV spectral
index $\alpha$ ($f_{\nu} \propto \nu^{\alpha}$) with redshift and only
a moderate trend with luminosity for the radio-loud objects.  However,
\citet{Scott 2004} concluded that the UV continuum from their
composite of 85 AGN spectra with $z \leq 0.67$ taken from \textit{Far
Ultraviolet Spectroscopic Explorer} (\textit{FUSE}) archival data, can
be characterized by a \textit{single} power law.  Given that the
\textit{FUSE} sample spectra have enhanced emission lines relative to
the \textit{HST} sample and have a median luminosity that is nearly an
order of magnitude lower, this suggests that the strength of any
continuum break may be luminosity dependent.  The results obtained
from the redshift and luminosity distributions of the {\it FUSE} and
{\it HST} AGN samples also suggest possible evolution of the UV SED
with redshift.

High S/N rest-frame UV/optical composites have also been constructed
from large samples in ground based optical surveys, such as the
composite of 688 optically-selected $0.2 < z < 3.0$ quasars from the
Large Bright Quasar Survey \citep{Francis 1991}, 657 radio-selected
$0.23 < z < 3.40$ quasars (with a larger fraction of low-redshift
objects) from the First Bright Quasar Survey \citep{Brotherton 2001},
and for 2200 optically-selected quasars spanning $0.04 < z < 4.80$
from the Sloan Digital Sky Survey \citep[SDSS;][]{Vanden Berk 2001}.
These studies have provided great insight into the detailed properties
of quasars longward of Ly $\alpha$, but optical spectra alone can only
probe the regions shortward of Ly $\alpha$ $\lambda1216$ at higher
redshifts ($z > 2$), where the density of Lyman-absorbing clouds is
high and significantly limits both the number of detections and the
S/N.

We investigate the results reported for \textit{FUSE} and \textit{HST}
UV composite SEDs by expanding the sample size by more than an order
of magnitude --- covering the full range of luminosity and redshift
spanned by both studies.  We also perform an empirical analysis of the
\textit{Galaxy Evolution Explorer} \citep[\textit{GALEX};][]{Morrissey
2005} astrometry and photometry that is crucial for any assessment of
the reliability of \textit{GALEX} positions and interpretations of the
resulting UV/optical colors and SEDs.  In \S\ref{sample}, we describe
the data sets used to find UV/optical candidates taken from the
\textit{GALEX} General Data Release 1 (GR1){\footnotemark} and the
Sloan Digital Sky Survey \citep[SDSS;][]{York 2000} Data Release 3
\citep[DR3;][]{Abazajian 2005}, as well as the description of the
construction of a clean and statistically significant photometric
sample of quasars.  An analysis of the UV/optical colors of the
photometric sample of \textit{GALEX}/SDSS matches, as well as the
resulting SEDs, are given in \S\ref{analysis}, and \S\ref{summary}
summarizes our results obtained from the clean sample and the
interpretation of the mean quasar SED luminosity and redshift
dependence.  Throughout this paper we assume a $\Lambda$-dominated
flat cosmology with $\Omega_{\Lambda} = 0.7, \Omega_m = 0.3,$ and
\textit{H}$_0$ = 70 km s$^{-1}$ Mpc$^{-1}$, consistent with the latest
WMAP results of \citet{Spergel 2006}.  In our study, none of the objects
in our UV/optical sample have been corrected for photometric contributions
from the host galaxy.  Unless otherwise indicated, we correct all
photometry for Galactic extinction using the extinction maps of
\citet{Schlegel 1998}.

\footnotetext{GR1 documentation can be accessed at
http://galex.stsci.edu/GR1/}

\section{THE DATA SET} 
\label{sample}

The inclusion of UV observations allows for (1) diagnostics that
can probe the evolution of the UV/optical SED with redshift, and (2)
observations of the rest-frame UV emission from lower redshift quasars
that are not heavily affected by the Ly $\alpha$ forest of absorption
\citep{Lynds 1971}, which one cannot accomplish with optical surveys
alone.  The accurate photometry ($\sim 0.02$ mag) and size (46,420
quasars) of the SDSS DR3 Quasar Catalog \citep{Schneider 2005} provides
a unique context for the interpretation of UV observations.

\subsection{\textit{GALEX}} 
\label{galex}

The \textit{GALEX} General Data Release 1 \citep[GR1;][] {Morrissey
2005, Martin 2005} includes observations covering nearly 8000 deg$^2$
of sky in four imaging surveys and two grism surveys as of December
2004. \textit{GALEX} is a NASA Explorer Mission that is performing an
All-Sky Imaging Survey (AIS) to $m_{AB} \approx 20.5$, a Medium
Imaging Survey (MIS) covering 1000 deg$^2$ to $m_{AB} \approx 23$, a
Deep Imaging Survey (DIS) covering 100 deg$^2$ to $m_{AB} \approx 25$,
and a Nearby Galaxy Survey (NGS) with a surface brightness limit of
$m_{AB} \approx 27.5$ arcsec$^{-2}$.  The observations are obtained in
two UV bands --- the far-ultraviolet (the FUV band) covering
$\sim1350-1750$ {\AA} and the near-ultraviolet (the NUV band) covering
$\sim1750-2750$ {\AA}.  The AB-magnitudes corresponding to the FUV and
NUV bands will hereafter be denoted by $f$ and $n$, respectively.  The
1\fdg25 field-of-view was designed to attain 4\arcsec-6{\arcsec}
resolution; currently the processing pipeline is achieving a FWHM of
6\farcs5 in the FUV band and 7\farcs2 in the NUV band \citep{Martin
2005}.  The large sample of objects provided by {\it GALEX}, along
with deep surveys in the optical and the near-IR, provides
multiwavelength coverage for a large number of galaxies, constraining
star-formation models and their evolution with cosmic time.  Lyman
absorption causes the flux of many optically identified sources to
quickly diminish with redshift at shorter wavelengths; however, the
extended coverage beyond the rest-frame Lyman limit at 912 {\AA} for
higher redshift objects has allowed for a number of UV counterparts to
optical SDSS sources to be identified.  For many objects,
\textit{GALEX} also allows a glimpse of the wavelength regime
shortward of Ly $\alpha$, and has permitted observations as blue as
$\sim 430$ {\AA} in the rest frame ($z < 2$), beyond which the Lyman
absorption becomes very strong.

\subsection{SDSS}
\label{sdss}

The SDSS will map $\sim 10^4$ deg$^2$ of the sky (mainly concentrated
at high Galactic latitudes) and obtain spectra for $\sim 10^6$
galaxies and $\sim 10^5$ quasars in five optical photometric bands
\citep[\textit{u,g,r,i,z;}][]{Fukugita 1996} to a depth of \textit{r}
$\sim 22$ \citep{York 2000} --- with photometric errors of less than
0.02 magnitudes \citep{Ivezic 2004}, and an astrometric accuracy
better than 0\farcs1 for \textit{r} $< 20.5$ \citep{Pier 2003}.  The
SDSS obtains imaging with a CCD mosaic camera \citep{Gunn 1998}
coupled to a dedicated 2.5m telescope \citep{Gunn 2006} at the Apache
Point Observatory.  The SDSS DR3 \citep{Abazajian 2005} contains more
than $10^8$ cataloged objects covering over 5000 deg$^2$ of sky, and
the Quasar Catalog includes observations of 46,420 type 1 quasars with
$M_{i} > -22$ and $z < 5.41$ \citep{Schneider 2005}.  In common with the
\textit{GALEX} photometric system, the \textit{u, g, r, i,} and
\textit{z} magnitudes are normalized to lie on the AB-system \citep{OG
1983, Smith 2002}.

The SDSS quasar survey is largely motivated by two science goals: (1)
to address the evolution of the quasar luminosity function with time
and (2) to examine the redshift dependence of quasar clustering.  The
SDSS quasar target selection algorithm \citep{Richards 2002} uses
point-spread function magnitudes corrected for Galactic extinction; it
involves targeting those objects that are unresolved in the FIRST
radio survey \citep{Becker 1995} and occupy distinct regions of
multi-dimensional color space outside of the stellar locus and the
regions occupied by nonactive galaxies --- optimized for the low- and
high-redshift objects and to detect quasars for $0 < z < 6$.  For
quasars with \textit{i} $ < 19.1$, $\approx$ 95\% are found by the
SDSS color-selection technique \citep{Vanden Berk 2005a}.  The DR3
Quasar Catalog (46,420 objects) consists of images taken between 1998
September 19 and 2003 May 1, as well as spectra obtained between 2000
March 5 and 2003 July 6.  The bright limit of the target selection
algorithm is \textit{i} = 15.0.

\subsection{Cross-Matching and Astrometry} 
\label{matching}

Previous work finding matches between \textit{GALEX} and the SDSS has
been implemented using \textit{GALEX} Early Release Observations (ERO)
for fields not restricted to quasars \citep[e.g.,][]{Agueros 2005},
for internal data releases \citep[e.g.,][]{Bianchi 2005, Seibert
2005}, and for small regions covered by {\em Spitzer} \citep{Richards
ph/06}.  We searched for UV counterparts to the 6371 SDSS DR3 quasars
covered by 204 \textit{GALEX} tiles in the \textit{GALEX} GR1 merged
object catalogs (hereafter referred to as the `quasar sample') --- using
an initial search radius $\theta$ = 7{\arcsec} \citep[roughly 1{\arcsec}
larger than the FWHM of the angular resolution limit in the NUV band;
for a more complete description see][] {Morrissey 2005}.  For a mean
SDSS quasar separation of $\sim$ 20{\arcmin} on the sky, the astrometry
of \textit{GALEX} (see below) and the SDSS is more than sufficient such
that the typical astrometric errors for a given source location will
be significantly smaller than the average quasar separation --- this
simplifies the matching routine significantly, since the probability of
more than one quasar lying within our search radius is extremely small.
In the case of more than one \textit{GALEX} UV source coincident with
an SDSS DR3 quasar to within 7\arcsec, the closest one was selected.
When overlapping \textit{GALEX} GR1 tiles each covered the same SDSS
quasar, we have selected the observation from the tile with the highest
effective exposure time ($t_{\rm eff}$) in the NUV band (and, in general,
correspondingly smaller photometric uncertainties for the same objects
due to higher S/N).

After matching the DR3 quasar sample with the GR1 catalogs, 5380
quasars (84.4\%) are detected in the NUV, while 3034 (47.6\%) are
detected in both the NUV and the FUV.  We also note that 38 quasars
(0.6\%) were detected only in the FUV (see discussion below).  There
were 953 (14.9\%) quasars with no GR1 counterparts; a statistical
analysis of the DR3 quasars with and without a UV detection is given
in \S\ref{statistics}.  The mean separation between a DR3 quasar and
the nearest {\it GALEX} UV source was $\approx$ 1\arcsec, with
$\approx 25\%$ of the quasars having more than one UV source within
7\arcsec.  If a specific tile has been observed multiple times, then
we use the coadded result; however, the {\it GALEX} source extractor can
and does identify UV sources unique to each epoch, so it is possible that
an appreciable fraction of the quasars with multiple UV sources within
7{\arcsec} are simply due to this phenomenon.  For the quasars with
multiple associated UV sources, the mean nearest separation is 0\farcs7,
while the mean separation of the next nearest UV source is 1\farcs4.

Our matches with only an FUV detection were rare --- with 18 sources
appearing on MIS tiles, 19 on AIS tiles, and a single source on a NGS
tile.  Diffuse reflections from bright stars located immediately
outside the {\it GALEX} detector field of view are not flagged as
image artifacts in the merged object catalogs, and can artificially
increase the number of false point source extractions for a given
field.  In addition, the source extractor of the {\it GALEX} image
processing pipeline will often fragment resolved objects into
individual UV sources, thereby contaminating search results with
multiple matches to the same optical counterpart{\footnotemark}.  For
the remainder of our analysis, we exclude the FUV-only matches to the
quasar sample, and given the small fraction of the total sample with
FUV only detections, we expect the remainder of our matches to be
negligibly contaminated by false detections.

\footnotetext{For additional GR1 documentation and a description of
data caveats see http://www.galex.caltech.edu/DATA/.}

\begin{figure}
\centering
\includegraphics[width=0.9\textwidth]{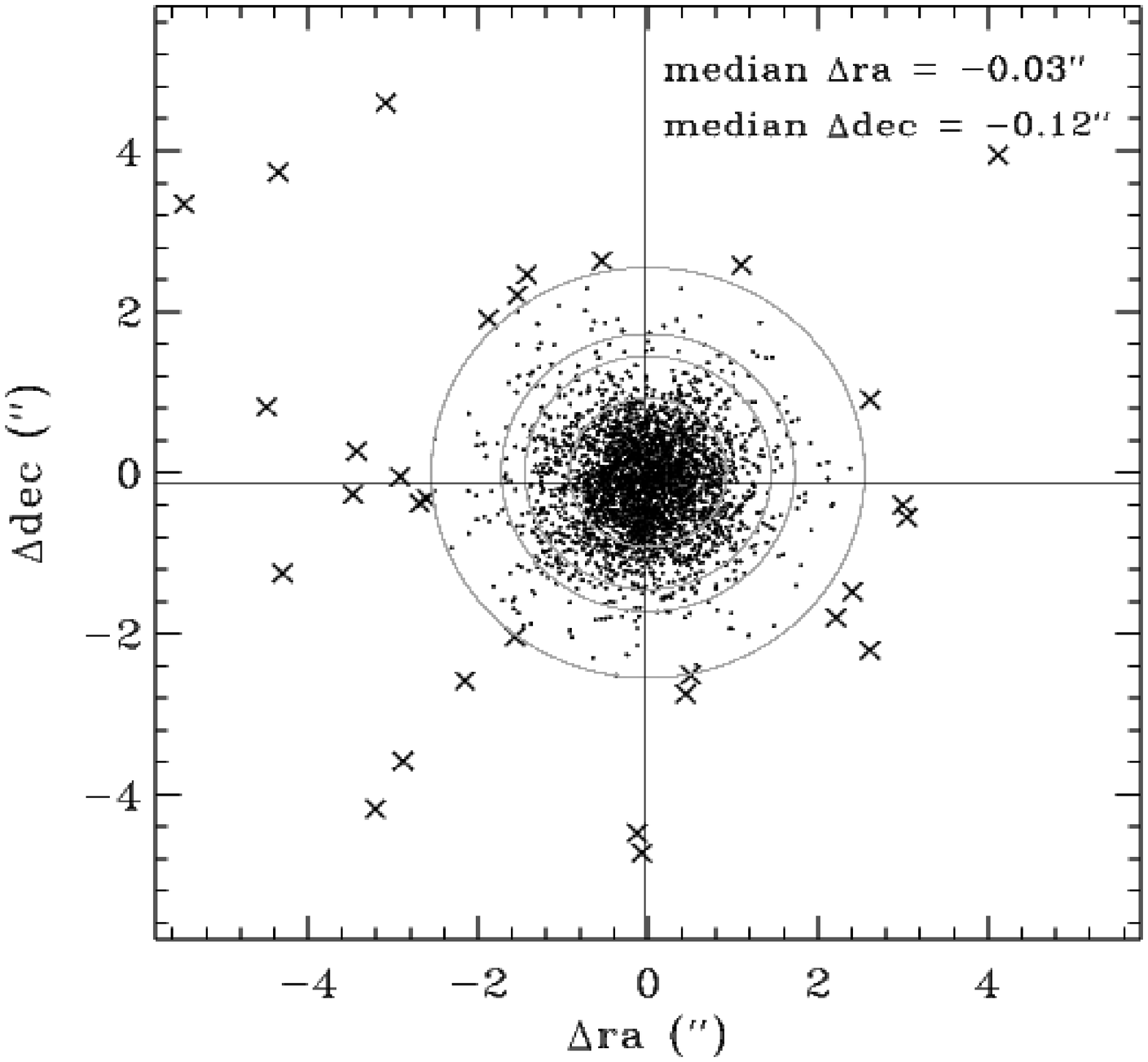}
\caption{Positional offsets for \textit{GALEX} GR1 matches in both UV
 bands (3034 quasars) to SDSS DR3 quasars.  The circles correspond to
 search radii $\theta$ of 0\farcs9, 1\farcs5, 1\farcs7, and 2\farcs6
 that recover 68\%, 90\%, 95\%, and 99\% of the matches, respectively.
 The x's identify the 30 objects found for 2\farcs6 $< \theta <$
 7\farcs0 that represent the outer 1\% of the detections.  The small
 points (3004 objects) represent the clean sample and are the quasars
 referred to and analyzed as `detections' throughout the remainder of
 this paper.  The median coordinate offsets ({\it GALEX} - SDSS) for
 all of the matches in both bands for right ascension and declination
 are -0\farcs03 and -0\farcs12, respectively.}
\label{coord_dispersion.fig} \end{figure}

Figure \ref{coord_dispersion.fig} illustrates the coordinate offsets
between SDSS and {\it GALEX} positions for all of the matches in both
UV bands.  These results show that $\approx 95$\% of our matches lie
within a matching radius $\theta$ = 1.7\arcsec, while $\approx 99$\%
are included if $\theta$ = 2.6\arcsec.  The median offsets in right
ascension and declination are -0\farcs03 and -0\farcs12 respectively.
Monte Carlo simulations were used to generate predicted distributions
of coordinate offsets between {\it GALEX} and SDSS positions, using a
set $10^4$ objects and assuming an rms $\sigma_{SDSS}$ = 0\farcs1 for
each SDSS coordinate.  A comparison of the simulation results with our
observed distribution of positional offsets shows that our statistics
are consistent with $\sigma_{GALEX}$ = 0.6-0.7{\arcsec} for each
\textit{GALEX} coordinate (the 0\farcs1 rms for the majority of SDSS
objects contributes negligibly when added in quadrature).  Our sample
is not significantly contaminated by false matches (there is no large
excess of objects at larger separations); although, as one might
expect, there are noticably more high-$\sigma$ matches than predicted
by our simulations, primarily because our Monte Carlo runs do not
take into account possible systematic errors affecting the {\it GALEX}
astrometry (e.g., the rate of false matches introduced by multiple source
extractions of resolved sources or image artifacts).  Our assessments
of the {\it GALEX} astrometry by making use of SDSS positions are fully
consistent with the astrometry derived from ground-based calibrations
conducted by the \textit{GALEX} collaboration\footnotemark.

For a completely random distribution of \textit{GALEX} UV sources in
the deepest (DIS) fields, the probability of any (star or galaxy) UV
source coincident with an SDSS quasar for $\theta$ = 7{\arcsec} is
$\approx 10\%$, while it is $\approx 1\%$ for $\theta$ = 2\farcs6.
Given that the majority ($> 99\%$) of the SDSS quasars are covered by
shallower tiles, we use 1\% as a liberal upper bound to our
contamination rate by non-quasar counterpart UV detections (either by
stars or image artifacts).  Accordingly, for the remainder of our
analysis and our discussion we will refer to the `detections' as those
matches with \textit{GALEX}/SDSS separations less than 2\farcs6 (5165
NUV detections and 3004 NUV+FUV detections).

\footnotetext{The {\it GALEX} collaboration website is
http://www.galex.caltech.edu/.}

\subsection{\textit{GALEX} Photometry}
\label{photometry}

In contrast to observations of relatively bright objects in the local
Universe (e.g., nearby resolved galaxies, stars, etc.) where the
photometric errors are usually small, photometric uncertainties of
relatively faint objects such as quasars in flux limited surveys are
no longer negligible and can largely determine the confidence of the
interpretation of the overall photometric properties of a sample.  UV
observations are more sensitive to line of sight extinction than
optical observations; thus quasars (being fainter and at higher
redshifts) are not as readily detected in the UV, despite emitting a
significant fraction of their bolometric luminosity in the UV region.
For the ones that are found in the UV, the photometric uncertainties
can become comparable to the detected flux itself, and one must
therefore know the distribution of errors well before forming
conclusions (e.g., the resulting UV colors and SEDs of
optically-selected quasars) in the context of the rest-frame emission.

The GR1 UV source extractor of the {\it GALEX} image processing
pipeline quotes photometric uncertainties for each object by assuming
the observations are Poisson noise limited (based on the number of
detected photons).  Therefore, the magnitude errors given in the GR1
object tables do not take into account additional sources of noise ---
including unknown variances of the detector background level and
flat-field maps, or any other systematic errors present in the data.
If one assumes the bulk of the stars observed in a {\it GALEX} field
are non-variable, then fields with multi-epoch observations are a
useful tool for analyzing the repeatability of the {\it GALEX} UV
photometry for the same objects, and one can use large numbers of
objects to empirically estimate the true photometric uncertainties.  In
\S\ref{analysis}, we incorporate the results of using this method
(described below) into our analysis of the UV colors of the quasars in
our sample.

\begin{figure}
\includegraphics[width=1.0\textwidth]{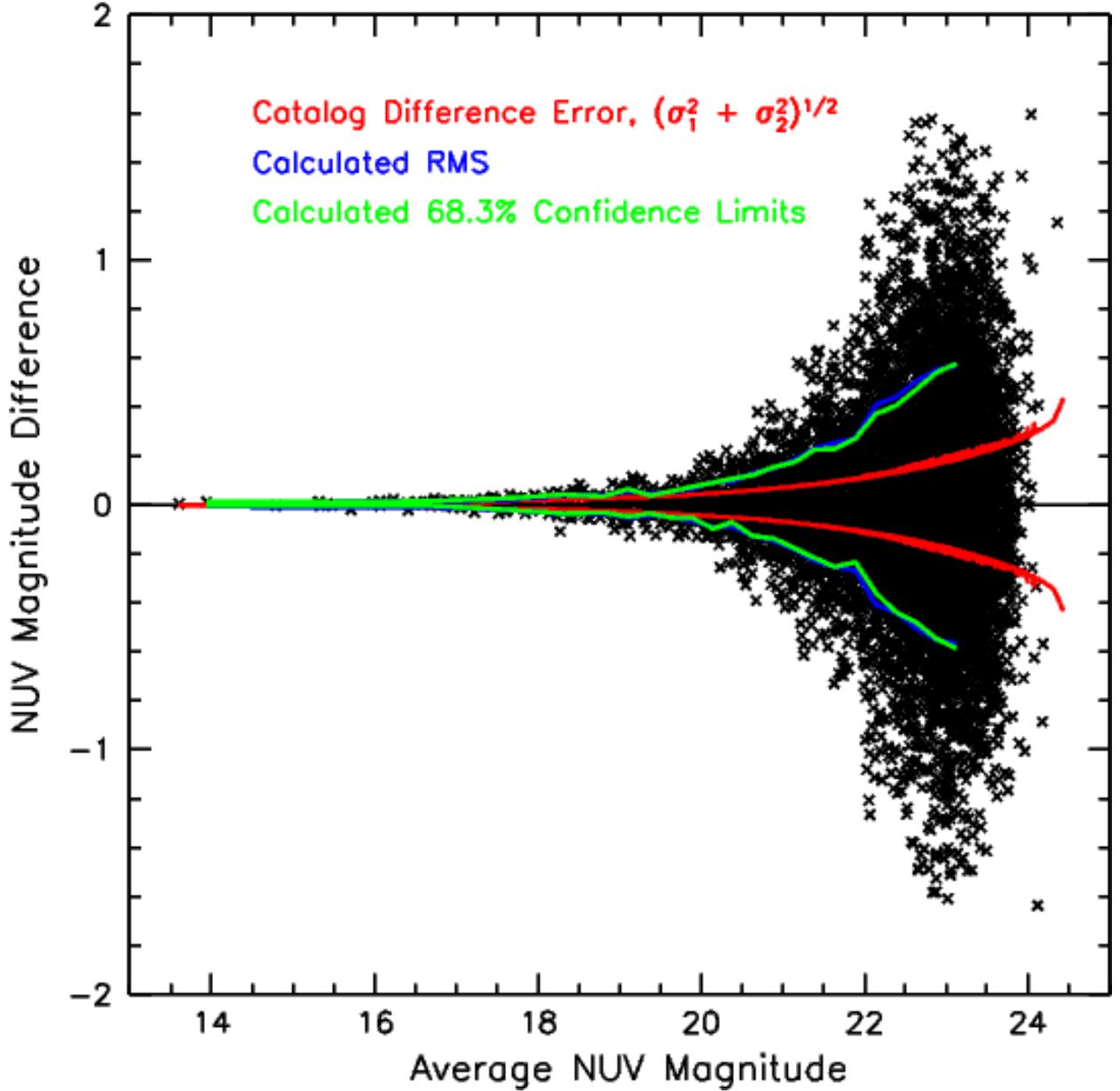}
\caption{A comparison of multi-epoch photometric measurements for the
 \textit{GALEX} Deep Imaging Survey (DIS).  A total of 6965 objects
 remain after outliers $> 3\sigma$ from the median value for the
 \textit{n}-magnitude differences have been removed --- likely to be
 spurious matches between both epochs of the same \textit{GALEX} tile.
 The red lines are the photometric errors taken from the GR1 merged
 object catalogs for each object after adding the errors from each
 epoch in quadrature, and are reflected about $\Delta n$ = 0; the blue
 curve represents the intrinsic rms dispersion calculated using
 appropriately sized \textit{n}-magnitude bins and is by definition
 symmetric about the \textit{n}-magnitude axis; the green curves are
 the calculated 68.3\% confidence limits for the objects.  The
 intrinsic rms very nearly overlaps our 68.3\% confidence limits ---
 indicating that the photometric errors are approximately Gaussian.
 The figure suggests that the true photometric errors are roughly a
 factor of $\sim 2$ larger than the published ones.}
\label{dis_nuverrors.fig}
\end{figure}

\begin{figure}
\includegraphics[width=1.0\textwidth]{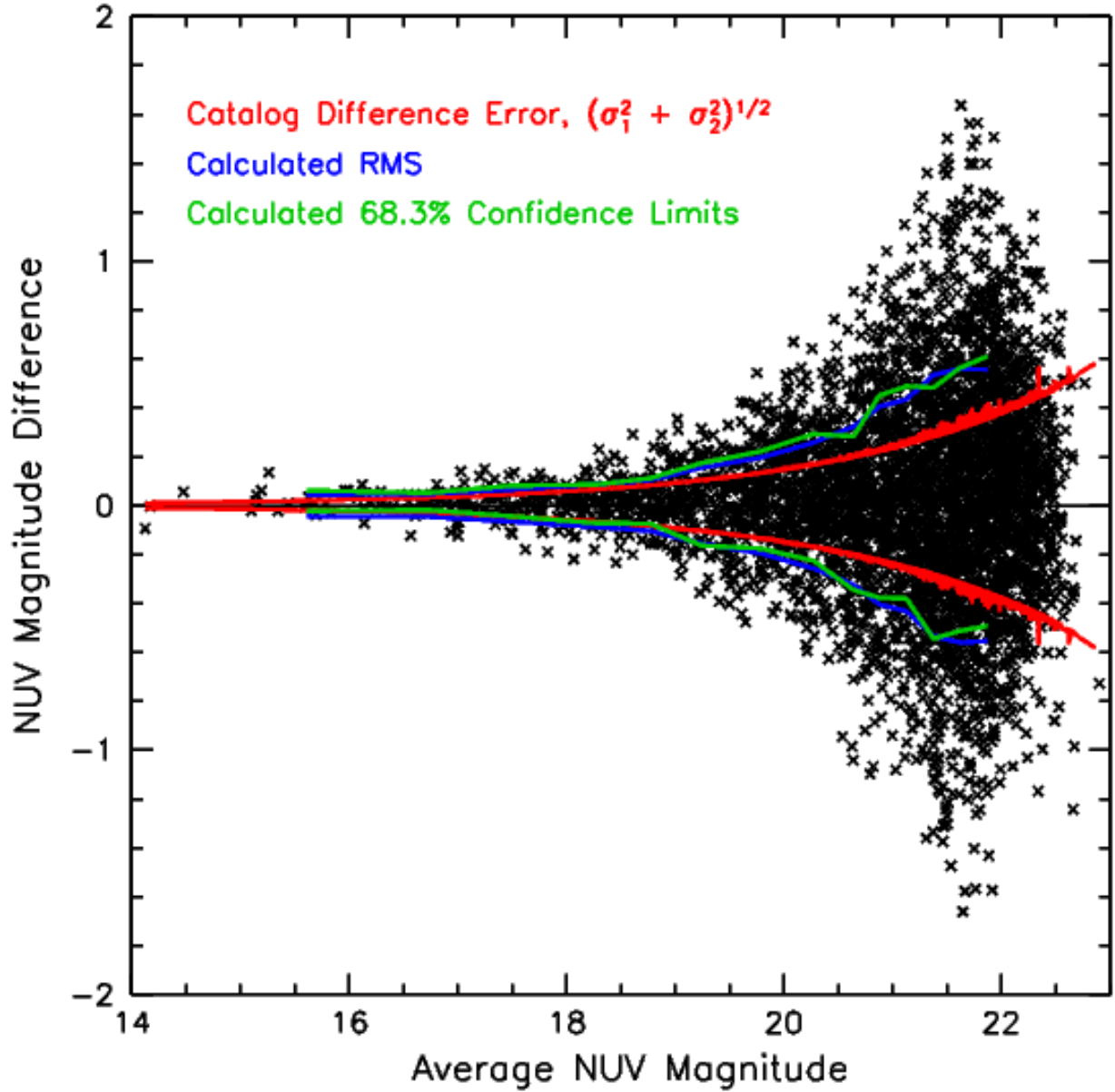}
\caption{Using the same notation as Figure \ref{dis_nuverrors.fig} for
 the \textit{GALEX} All-Sky Imaging Survey (AIS) observed at multiple
 epochs.  3966 objects remain after outliers $> 3\sigma$ from the
 median value for the \textit{n}-magnitude differences were removed.}
\label{ais_nuverrors.fig}
\end{figure}

Using multi-epoch \textit{GALEX} tiles from the \textit{Spitzer First
Look Survey} area with similar exposure times taken from the Deep
Imaging Survey, we matched 7090 NUV detections using a search radius
of 7{\arcsec} ($\sim$ FWHM in the NUV band) and restricted the
matching out to 95\% of the tile radius, thus minimizing the
contamination by image artifacts and multiple source extractions of
diffuse out-of-field bright stars that can dominate the sources near
the edge of the fields.  After removing matches with magnitude
differences $> 3\sigma$ from the median of each magnitude bin, 6965
(98.2\%) objects remain.  Figure \ref{dis_nuverrors.fig} shows the NUV
magnitude differences ($\Delta$NUV) as a function of NUV magnitude,
with the red curves showing the errors given in the GR1 tables after
adding the errors in quadrature for each epoch (and reflected about
$\Delta$NUV = 0).  The green and blue curves are the rms and 68.3\%
confidence limits respectively, which were calculated in NUV magnitude
bins.  There is good agreement between the rms and confidence limits,
indicating that the errors in the NUV photometry are approximately
Gaussian distributed at the locations of each magnitude bin.  A
similar analysis was also performed for repeated FUV observations, as
well as for both the NUV and FUV bands in the All-Sky Imaging Survey
(e.g., Figure \ref{ais_nuverrors.fig}).

\begin{figure}
\includegraphics[width=0.9\textwidth]{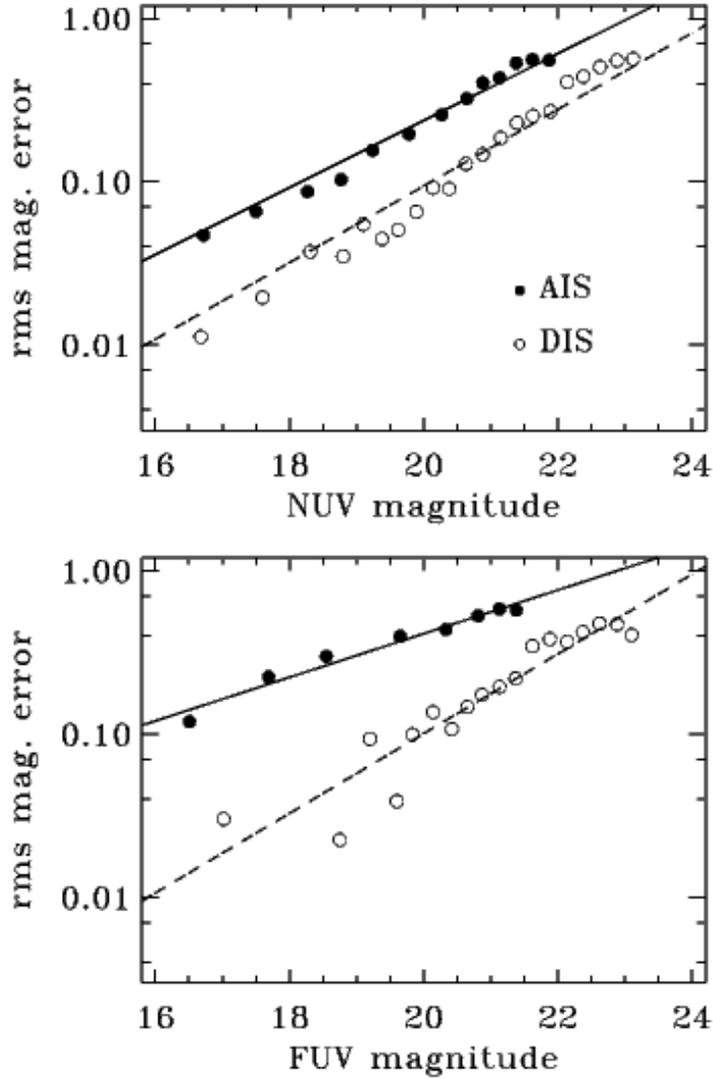}
\caption{The binned rms magnitude errors as a function of magnitude
 from matches on multi-epoch \textit{GALEX} tiles for the NUV
 (\textit{top}) and the FUV (\textit{bottom}).  The solid and dashed
 lines represent the least-$\chi^2$ fits to the binned distribution of
 empirically determined photometric errors, for AIS and DIS fields
 respectively.}
\label{photrms.fig}
\end{figure}

\begin{figure}
\includegraphics[width=1.0\textwidth]{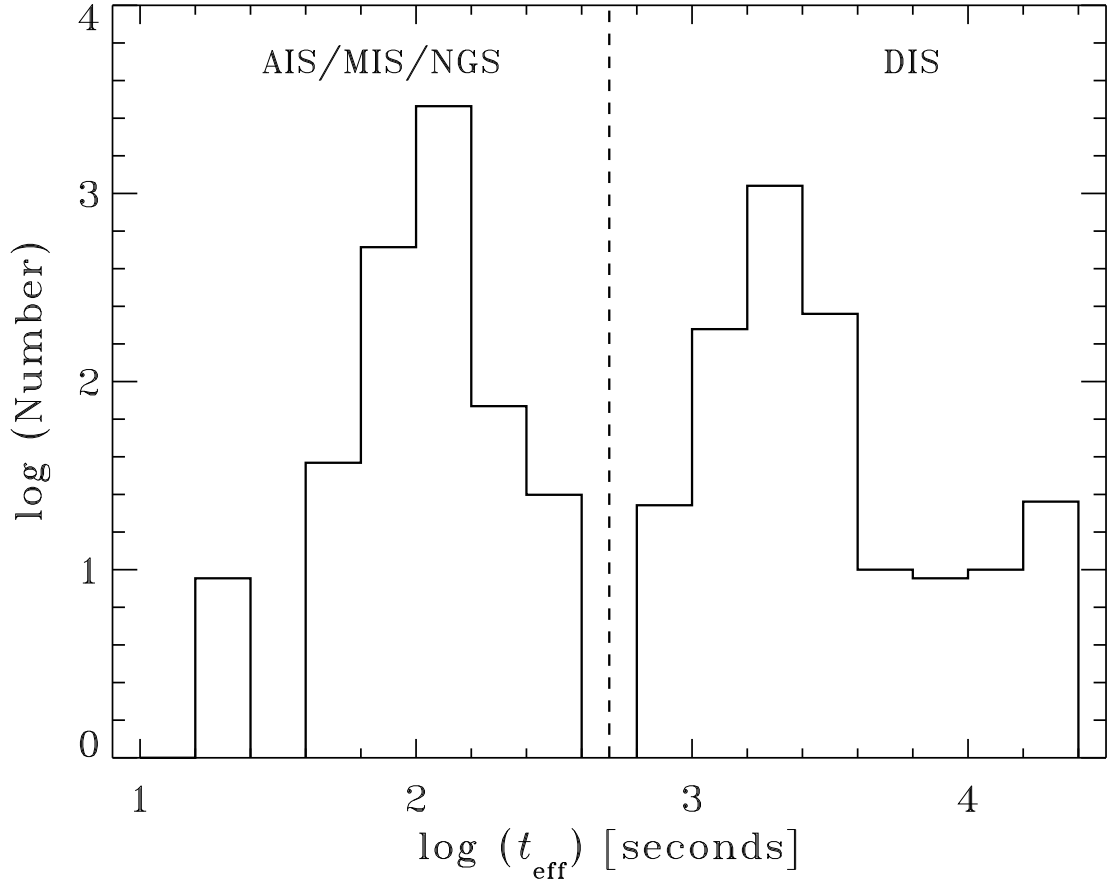}
\caption{Histogram showing the overall distribution of NUV effective
 exposure times for the DR3 quasars detected by {\it GALEX}.  For the
 objects to the left of the empirical division at log $(t_{\rm eff})$
 = 2.7 (\textit{dashed line}), we use the AIS empirical curves of
 Figure \ref{photrms.fig}, and for objects to the right of this line,
 we use the DIS empirical curves to compute our estimates of the true
 NUV and FUV photometric errors for all of the UV detections in our
 sample.}
\label{teff_hist.fig}
\end{figure}

Figure \ref{photrms.fig} shows the least-squares fits to the binned
NUV and FUV magnitude rms as a function of magnitude.  Open and closed
symbols are calculated from the DIS and AIS results, respectively,
as discussed above.  Due to the mean exposure time for an AIS tile
being roughly an order of magnitude lower than that for a DIS tile,
the AIS empirical error best-fit curves lie above those for the DIS
due to lower S/N.  The qualitatively larger error and fewer bins in
the FUV results from the lower density of FUV sources on the sky above
the threshold flux and the lower sensitivity of the FUV band relative
to the NUV band.  Figure \ref{teff_hist.fig} shows the distribution of
effective exposure times for all of the quasar matches in both UV bands.
Given the bimodal distribution, we divide the sample at an effective NUV
exposure time of log $(t_{\rm eff}) = 2.7$ --- for our
empirical estimates of the true photometric uncertainty, we employ the AIS
curves for detections with log $(t_{\rm eff})$ $< 2.6$ and the DIS curves
for log $(t_{\rm eff})$ $> 2.6$ (see Figure \ref{photrms.fig}).  If we
define $a$ and $b$ to be the slope and y-intercept of the overplotted
curves respectively, then the empirical photometric error ($\Delta m$)
is given by
\begin{eqnarray}
\log{\Delta m} &=& am + b,
\label{emp_error}
\end{eqnarray}
where $m$ is the quasar NUV/FUV magnitude.  Table \ref{phot_err.tab} gives
our best-fit values for $a$ and $b$, which we have used to determine the
empirical photometric errors given in our table of matching data (\S2.5).

From the number of repeated NUV detections across multiple epochs for
a DIS field, we estimate an $ \sim 8$\% probability that a quasar from
the SDSS DR3 will have have a false match in the NUV band for a search
radius of 7\arcsec, while this drops to $<$ 1\% if the search radius
is limited to 2.6{\arcsec} --- this justifies the selection process
discussed previously in \S\ref{matching}.  In addition, we extend this
contamination rate to the entire quasar sample, given that these
calculations are based on the deepest {\it GALEX} fields with the
highest density of UV sources.  The results of the empirical analysis
of the {\it GALEX} NUV and FUV photometry clearly show the true
photometric error rises above the Poissonian errors quoted in the GR1
tables (e.g., Figure \ref{dis_nuverrors.fig}), where for the NUV band
quasar detections (median $n \sim 19.8$), the photometric errors
(assuming the stars to be non-variable) are already in excess of $\sim
5$\% --- further justifying this analysis as necessary for constraining
the behavior of the quasar UV colors with redshift, the relative
colors (\S\ref{colorzed}), and the UV/optical SEDs (\S\ref{seds}).

\subsection{Data Table}
\label{table}

Our database (see Table \ref{sample_table.tab}) of \textit{GALEX} GR1
matches to SDSS DR3 quasars contains all of the parameters listed in
Table \ref{table.tab} (and described below) for each of the 6371 SDSS
DR3 quasars covered by GR1 tiles.  Each entry consists of selected
parameters from both the GR1 catalogs and the SDSS DR3 quasar catalog,
as well as our measured quantities.  Full versions of the
\textit{GALEX} GR1 tables are publicly available at
{http://galex.stsci.edu/GR1/} and the DR3 quasar catalog is available
via the SDSS public website at {http://www.sdss.org}.  Of the 6371
quasars in the table, 5380 (84.4\%) are NUV detections and 3034
(47.6\%) are NUV+FUV detections, found by matching coordinates and
before any quality cuts are made (see \S\ref{matching}).  The table
also includes 38 (0.6\%) FUV only detections --- these objects are
likely contamination by false matches (a mean separation of $\sim
2.2$\arcsec) and we therefore do not refer to them as `detections'
for the purposes of our analysis.

For the DR3 sample quasars without an NUV and/or FUV source within
7{\arcsec}, the entries in columns 33 - 34 are the NUV $t_{\rm eff}$
and/or FUV $t_{\rm eff}$ taken from the tile that covered the quasar ---
the GR1 tables do not give upper limits in magnitudes.  In principle,
the relative $t_{\rm eff}$ would be the only parameter necessary to
calculate upper limits for a given tile, but the non-uniform response of
the {\it GALEX} detectors as well as changing UV background levels would
necessitate calculating upper limits on a tile-by-tile basis.  Future {\it
GALEX} data releases may enable a more straightforward calculation of
upper limits for each optically-selected quasar without a UV detection.

Description of the table parameters:

1. The SDSS DR3 object designation from the quasar catalog in the
   format SDSSJhhmmss.ss+ddmmss.s, where `SDSSJ' is omitted.

2. The \textit{GALEX} GR1 object designation from the merged object
   catalog.  The same object identified on multiple \textit{GALEX}
   tiles can have more than one object id.  Thus, one should always
   use object coordinates and not object designations to identify
   individual objects as such.  DR3 quasars without a UV detection
   (see \S\ref{matching}) will have a `-999' entry.

3-4. The object coordinates (right ascension and declination; J2000.0)
   from the SDSS DR3 quasar catalog, given in decimal degrees.  For a
   significant majority of the objects, the SDSS astrometric accuracy
   of each coordinate is 0\farcs1, with the largest errors expected
   not to exceed 0\farcs2 \citep[see][]{Pier 2003}.

5-6. The object coordinates (right ascension and declination; J2000.0)
   from the \textit{GALEX} GR1 merged object catalogs, given in
   decimal degrees.  Our empirical analysis of \S\ref{matching}
   demonstrates the derived rms of $\sim$ 0\farcs6 - 0\farcs7 for each
   \textit{GALEX} coordinate.  DR3 quasars without any UV detection
   will have a `-999' entry for both coordinates.

7. The coordinate offsets between the {\it GALEX} and SDSS positions
   ({\it GALEX} - SDSS), given in arcseconds.  Objects with
   separations $>$ 7{\arcsec} or non-detections have a `-999' entry

8. The quasar redshifts from the DR3.  For any given object, the
   estimated redshift uncertainty is $< 0.01$, except for Broad
   Absorption Line (BAL) quasars \citep[e.g.,][]{Trump 2006}.

9-18. The DR3 PSF magnitudes and errors for all of the quasars (no
   Galactic extinction corrections applied), given in AB-magnitudes
   from the five SDSS photometric (\textit{u, g, r, i,} and
   \textit{z}) bands.  The magnitudes are those derived from BEST
   photometry (or TARGET if BEST is unavailable; see \citet{Schneider 2005}
   and references therein).

19-24. The {\it GALEX} PSF magnitudes and errors (both as quoted in
   the GR1 object catalogs and from our empirical analysis of the
   photometry), in units of AB-magnitude from both UV photometric
   bands.  The effective wavelengths of the FUV and NUV channels are
   1528 {\AA} and 2271 {\AA}, respectively \citep{Morrissey 2005}.

25. The Galactic extinction in the SDSS-\textit{u} band ($A_u$)
    derived from \citet{Schlegel 1998}.  The extinction in the
    remaining SDSS bands can be expressed as $A_g, A_r, A_i,$ and
    $A_z$ with values corresponding to 0.736, 0.534, 0.405, and 0.287
    times $A_u$.  For the {\it GALEX} $f$ and $n$ bands, the
    corrections $A_f$ and $A_n$ are 1.584 and 1.684 times $A_u$,
    respectively \citep{Wyder 2005}.  All correction factors assume an
    $R_V = 3.1$ absorbing medium.

26. Log $N_H$ represents the logarithm of the neutral Hydrogen column
    density along our Galactic line of sight to each quasar,
    interpolated via 21 cm data (also see \citet{Schneider 2005} and
    references therein).

27. The absolute \textit{i}-band magnitude --- calculated after
    correcting for Galactic extinction and assuming a power-law
    continuum index of $\alpha = -0.5$ ($f_{\nu} \propto
    \nu^{\alpha}$), and a $\Lambda$-dominated flat cosmology with
    $\Omega_{\Lambda} = 0.7, \Omega_m = 0.3,$ and \textit{H}$_0$ = 70
    km s$^{-1}$ Mpc$^{-1}$ (same for this paper).

28. The logarithmic monochromatic quasar luminosity ($\nu$L$_{\nu}$)
    computed at 2200 {\AA} (rest-frame).  A `-999' entry is given if
    there is no wavelength coverage.

29. The same as column 28, but for 5100 {\AA} --- only computed for the
    quasars where 5100 {\AA} has not been redshifted out of the
    SDSS-$z$ band.  A `-999' entry is given if there is no wavelength
    coverage.

30-31. The $\Delta(g-i)$ and $\Delta(f-n)$ relative colors
    (\S\ref{colorzed}), defined as the 
    $g-i$ and $f-n$ colors minus the corresponding median colors at the
    redshift of the quasar, given in units of AB-magnitude.

32. Object distance from \textit{GALEX} field center in decimal
    degrees.  Note that for the case of objects with multi-epoch
    observations across multiple \textit{GALEX} fields, the entry from
    the tile with the highest effective NUV exposure time
    (`NUV\_WEIGHT' in the GR1 catalog) is used (see \S\ref{matching}).

33. The \textit{GALEX} NUV effective exposure time in seconds.  The
    non-detections have a `-999' entry.  For the 991 DR3 quasars
    covered by a \textit{GALEX} tile for which no NUV source was found
    within 7{\arcsec} (a `-999' entry for columns 22-24), the NUV
    effective exposure time for the next nearest near-UV source to the
    DR3 quasar is given in seconds.

34. The \textit{GALEX} FUV effective exposure time in seconds.  The
    non-detections have a `-999' entry.  Of the 3072 FUV detections,
    38 are {\it only} detected in the FUV (a `-999' entry for columns
    19-21), and are not analyzed as true matches in this study (see
    the discussion in \S\ref{matching}).

35. The \textit{GALEX} survey flag.  The possible values are integers
    from 1-4 that specify the survey from which the UV identification
    of the SDSS quasar was drawn (4 = AIS, 3 = NGS, 2 = MIS, and 1 =
    DIS; also see \S\ref{galex} for the data/survey descriptions).

\section{ANALYSIS} 
\label{analysis}

\subsection{Matching Statistics} 
\label{statistics}

\begin{figure}
\includegraphics[width=1.0\textwidth]{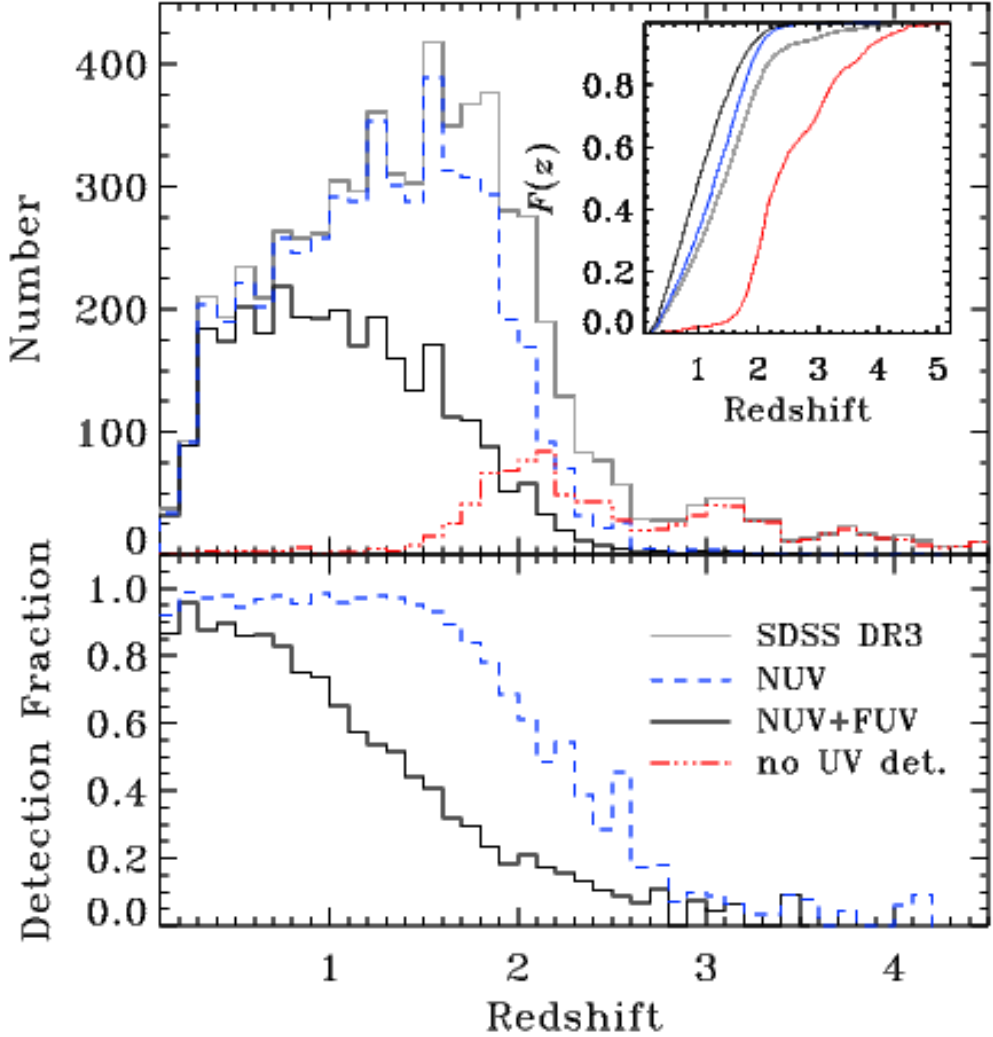}
\caption{Quasar redshift distributions. (\textit{top}) --- The solid
 black line shows the binned distribution of quasar redshifts ($0.12 <
 z < 3.41$; median 1.02) for the 3004 \textit{GALEX}/SDSS detections
 in both UV bands.  The same distribution for the DR3 quasars covered
 by all \textit{GALEX} GR1 fields ($0.12 < z < 5.20$; median 1.45)
 is the solid gray line.  The blue dashed line is for all of the DR3
 quasars with an NUV detection, and the red line represents quasars
 without a UV detection.  The local minima in the DR3 distribution (at
 $z \sim 2.7$ and at $z \sim 3.4$) correspond to redshifts where the
 DR3 target selection algorithm has lower efficiency, as is discussed
 by \citet{Schneider 2005} and \citet{Richards 2006}.  The inset shows
 the cumulative distribution functions for each subsample, using the
 same color scheme.  (\textit{bottom}) --- The detection fraction for
 the quasars found in both \textit{GALEX} bands (black solid line) as
 well as for the NUV (blue dashed line).  In the NUV, nearly all of the
 DR3 sample quasars out to $z \sim 1.7$ are detected by {\it GALEX}.}
\label{z_hist.fig} 
\end{figure}

Both intrinsic differences among quasars (e.g., due to reddening by dust
within the host galaxy) and observational effects (e.g., the Ly $\alpha$
forest entering the FUV band at $z \sim 0.1$ and the NUV band at $z \sim
0.5$) will affect the detection probability in UV observations.  The top
panel of Figure \ref{z_hist.fig} illustrates the redshift distribution
of the DR3 sample (gray line), the NUV detections (blue dashed-line),
the NUV+FUV (black solid-line) detections, and the non-detections
(red line), using redshift bins of width $\Delta z = 0.1$.  In the
DR3 sample, the local minima seen at redshifts $z > 2.5$ correspond to
where the SDSS star-quasar separation is less efficient ($z \sim 2.7$
and $z \sim 3.4$), also where quasars cross into the stellar locus in
multi-dimensional color-color space \citep{Richards 2002}.  The bottom
panel shows the respective fraction of DR3 quasars detected for the
same distributions using the same color scheme.  The NUV+FUV detection
fraction begins to noticeably decrease as soon as Ly $\alpha$ leaves the
FUV band ($z \sim 0.4$).  Note that in the NUV band, we recover nearly
all of the DR3 quasars until $z \sim 1.4$ (also where the Ly $\alpha$
forest is extinguishing flux in the entire {\it GALEX} bandpass), with the
detection fraction dropping to $\sim$ 50\% by $z \sim 2$.  The resulting
UV/optical colors from these matches are discussed in \S\ref{colorcolor}.

\begin{figure}
\centering
\includegraphics[width=0.95\textwidth]{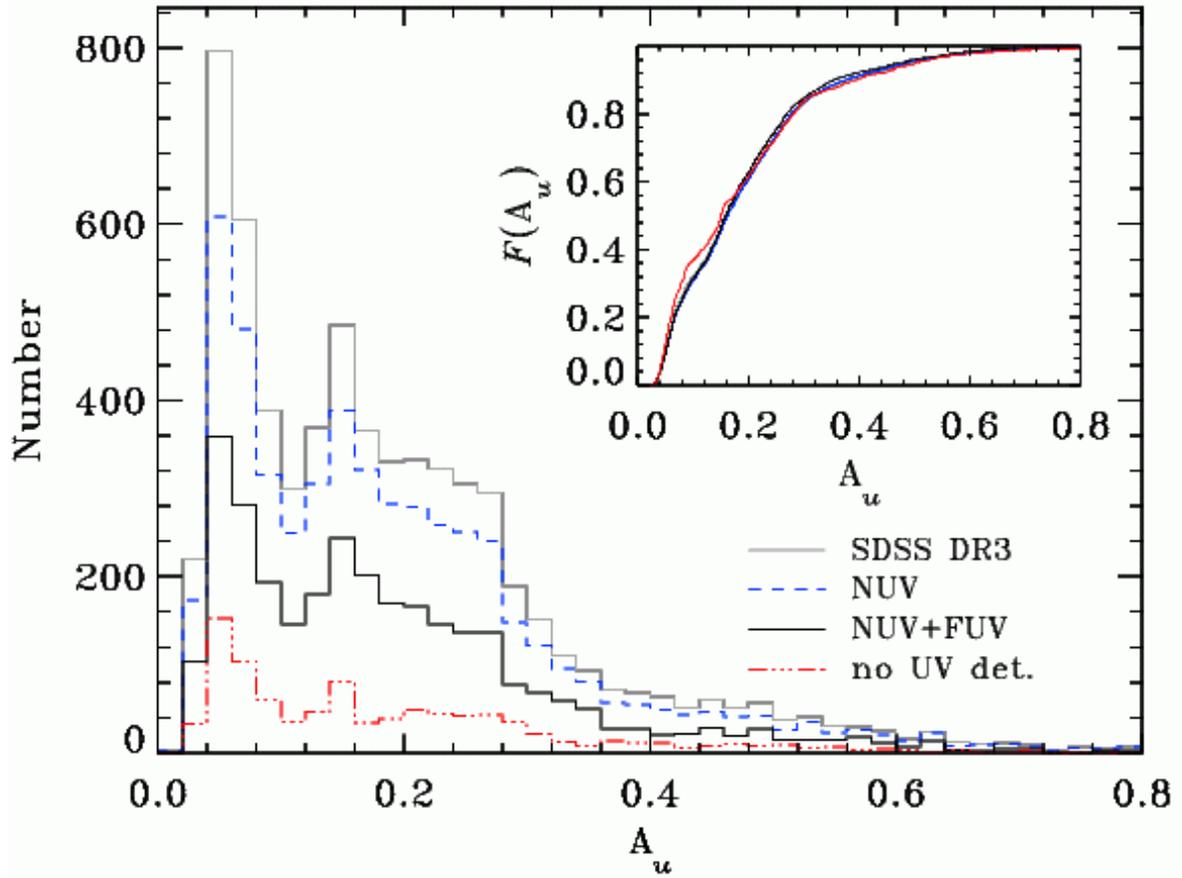}
\caption{Using the same notation as Figure \ref{z_hist.fig}, the
 distributions of Galactic extinction in the SDSS $u$-band using bins
 of width $\Delta A_u = 0.025$.}
\label{ks_au.fig}
\end{figure}

\begin{figure}
\centering
\includegraphics[width=0.95\textwidth]{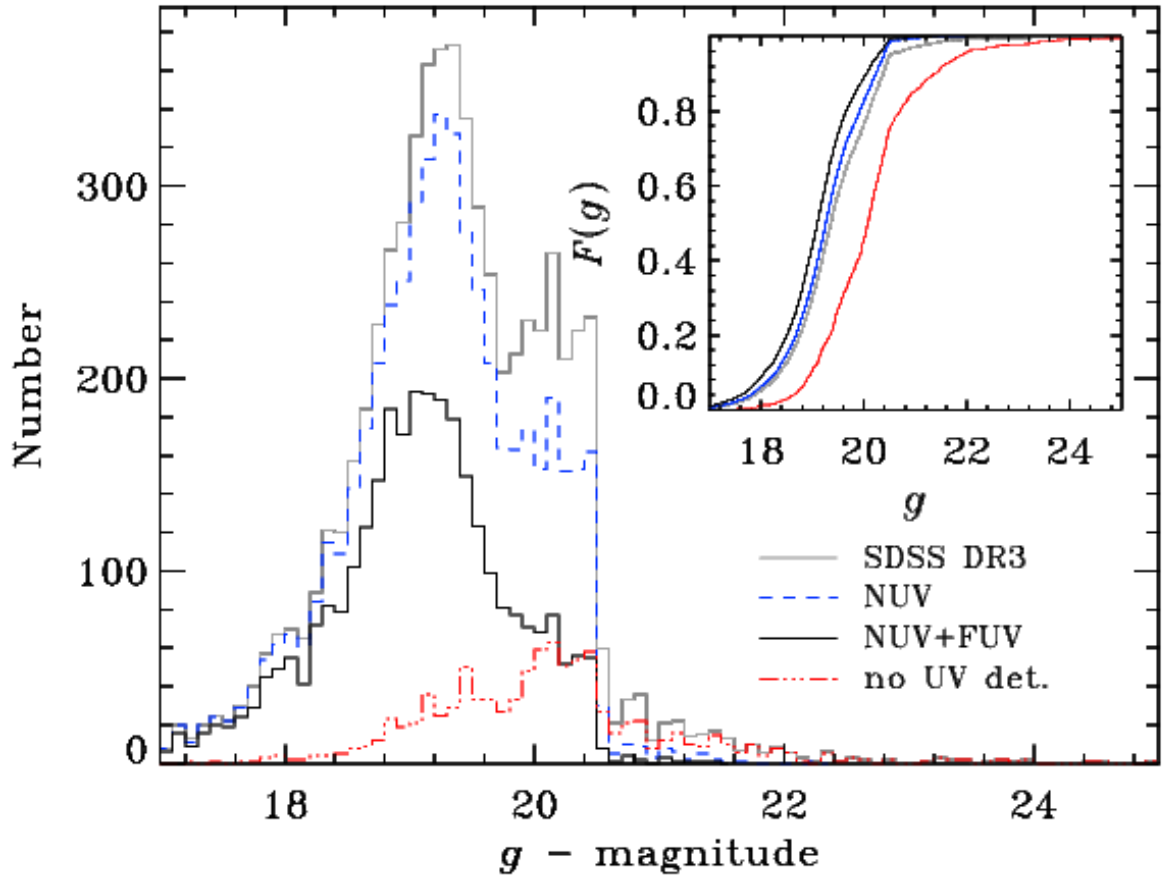}
\caption{Using the same notation as Figure \ref{z_hist.fig}, here for
 the optical $g$-magnitude distributions (no Galactic extinction
 corrections have been applied here --- see \S{\ref{analysis}}), using
 magnitude bins of width $\Delta g = 0.1$.}
\label{ks_g.fig}
\end{figure}

\begin{figure}
\centering \includegraphics[width=0.95\textwidth]{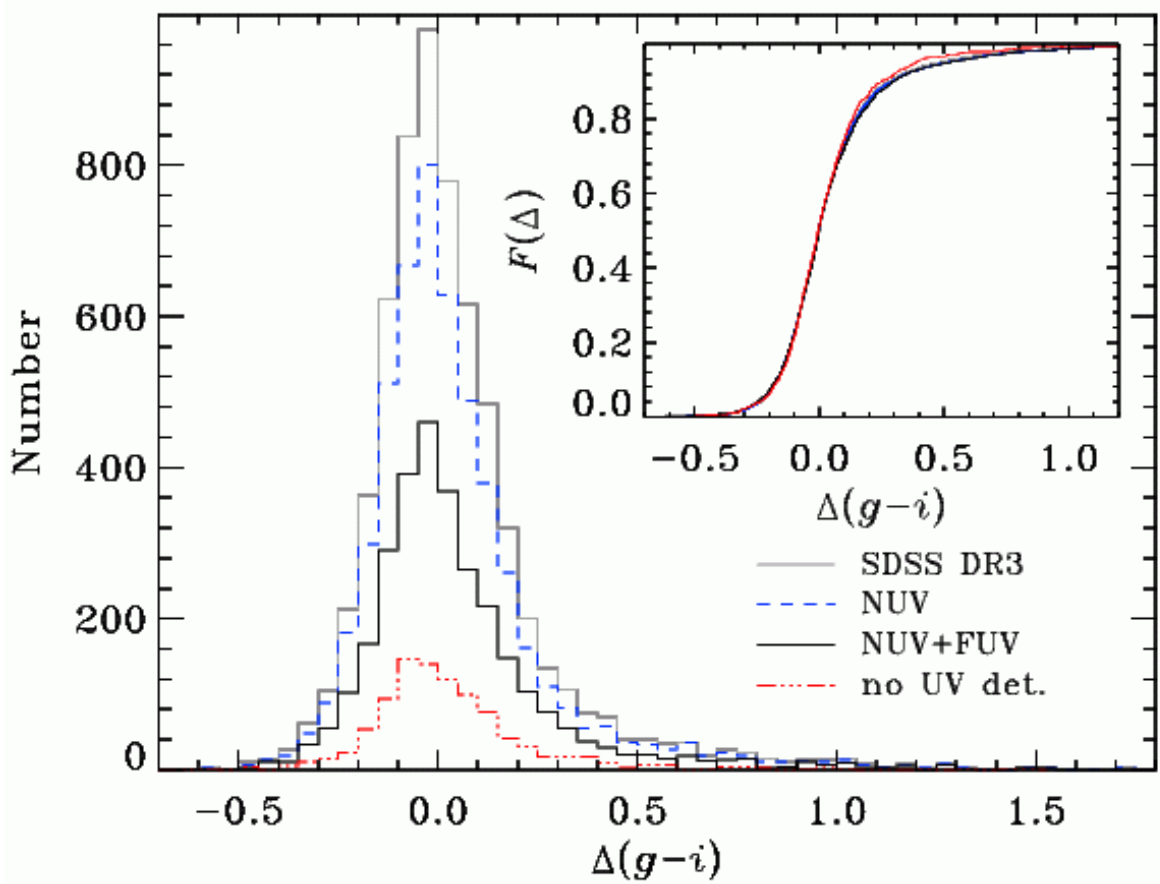}
\caption{Using the same notation as Figure \ref{z_hist.fig}, the
 relative color $\Delta (g-i)$, which is defined as the difference
 between the $g-i$ color of the quasar and the median $g-i$ color at
 its redshift.  The bins have a width of 0.1.}
\label{ks_dgi.fig} \end{figure}

\begin{figure}
\centering
\includegraphics[width=1.0\textwidth]{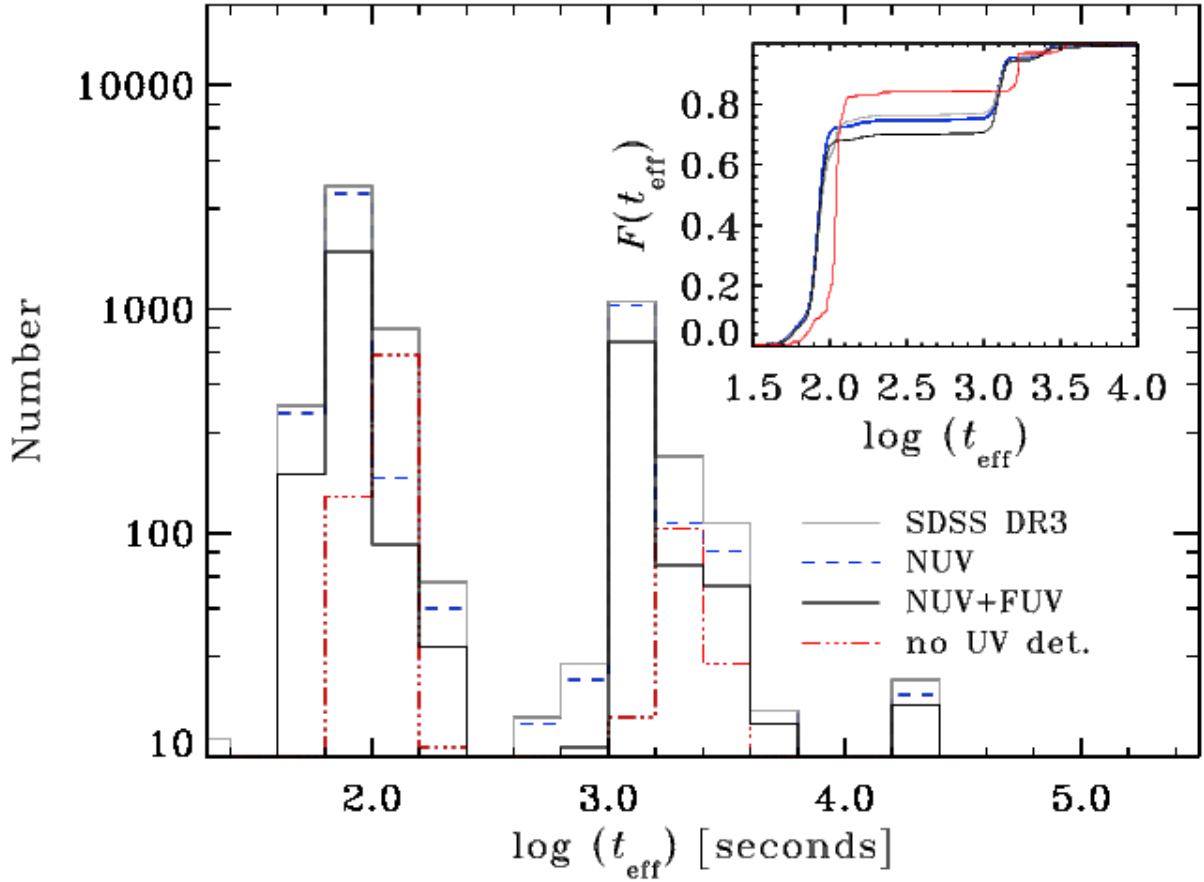}
\caption{Using the same notation as Figure \ref{z_hist.fig}, here for
 the NUV effective exposure time ($t_{\rm eff}$) as
 defined in \S\ref{matching}, using bins of width dlog
 $t_{\rm eff}$ = 0.2.}
\label{ks_teff.fig}
\end{figure}

For our statistical tests of the matching results, we analyzed the UV
detections and non-detections by dividing our dataset into four
subsamples: (1) the parent sample of DR3 quasars, (2) those with an
NUV detection, (3) those with an NUV+FUV detection, and (4) those with
no UV detection.  The FUV-only detections were not analyzed as a
separate subsample (see discussion in \S\ref{matching}).  Figures
\ref{z_hist.fig} - \ref{ks_teff.fig} display the distributions of each
as functions of redshift, Galactic extinction in the $u$ band ($A_u$),
$g$-magnitude, relative optical color $\Delta(g-i)$, as well as NUV
effective exposure time $(t_{\rm eff})$.  The inset of each histogram plot
also displays the cumulative distribution functions for each
parameter.  Table \ref{ks_table.tab} summarizes the results of
performing the two-sided Kolmogrov-Smirnov (KS) test on each subsample
relative to the parent DR3 quasar sample.  The results of the KS tests
indicate that brightness ($g$-magnitude) and redshift are the primary
explanations for non-detections in the UV.  Gas and dust along our line of
sight preferentially attenuates the flux most strongly in the UV bands,
which is consistent with the KS test results that show Galactic extinction
($A_u$) is likely contributing to the lower number of detections in
the FUV band.   Although the KS statistics would suggest $t_{\rm eff}$
as another explanation for the non-detections, an inspection of Figure
\ref{ks_teff.fig} reveals the subsample of non-detections appears to
be more heavily weighted toward larger $(t_{\rm eff})$, which is in
the opposite sense of what one would expect for the quasars with no
UV detection.

\subsection{Color-Color Relations}
\label{colorcolor}

\begin{figure}
\centering 
\includegraphics[width=1.0\textwidth]{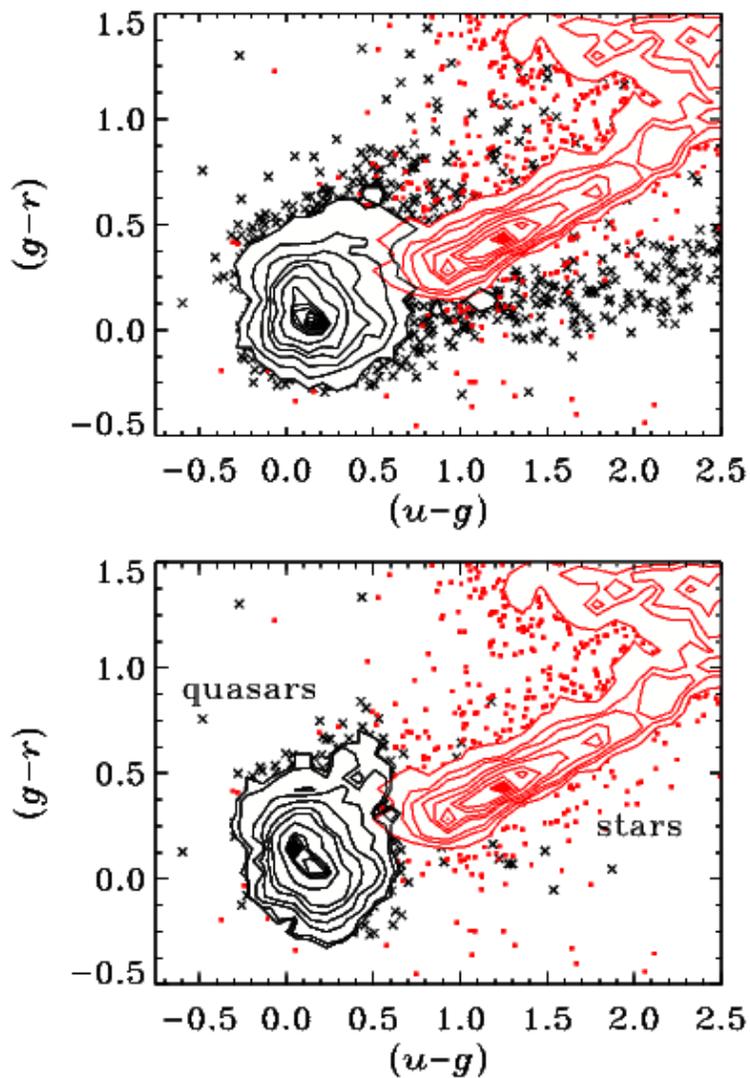}
\caption{(\textit{top}) The black x's and contours show the
 distribution of colors for 6371 SDSS DR3 quasars covered by
 \textit{GALEX} GR1 fields.  The red points and contours are stars
 from the DR4 database.  (\textit{bottom}) Identical to the top panel,
 but only the quasars detected in both \textit{GALEX} UV bands are
 shown along with the same set of SDSS stars.  All objects have been
 corrected for Galactic extinction.} 
\label{dr3_colors.fig}
\end{figure}

\begin{figure}
\centering
\includegraphics[width=1.0\textwidth]{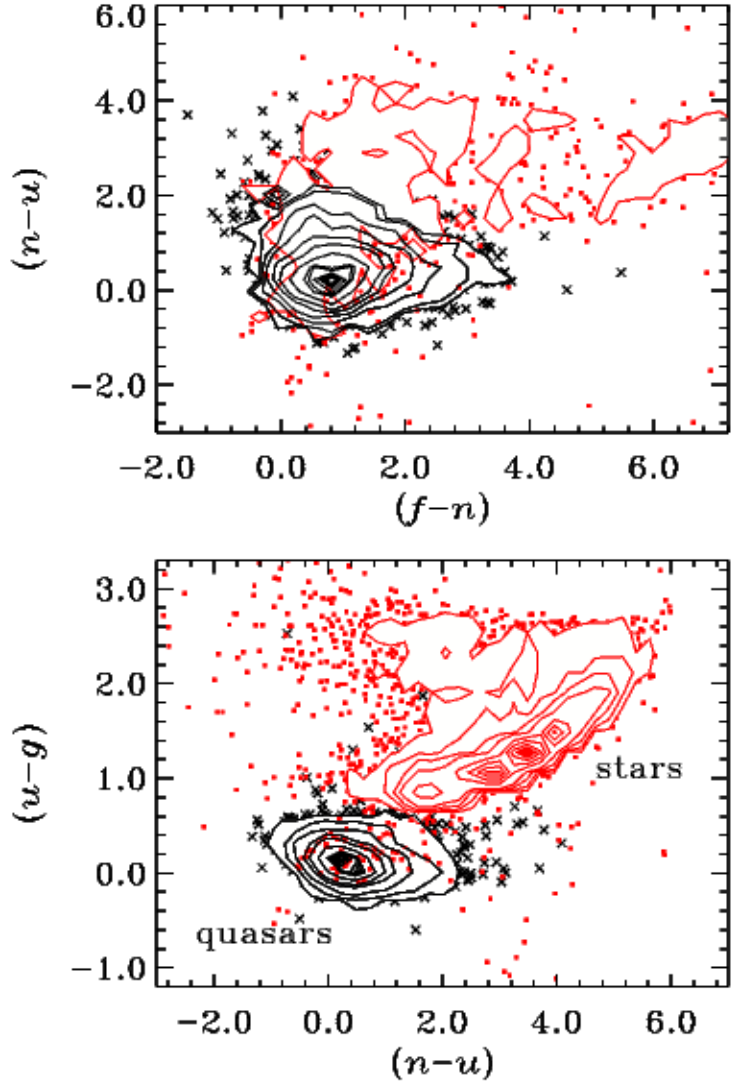}
\caption{The black points and contours show the distribution of colors
 for 3004 quasars detected in both \textit{GALEX} bands.  The red
 points are stars from GR1 DIS fields matched to DR4 stars ($g <
 20.5$).  All objects have been corrected for Galactic
 extinction.} 
\label{uvopt_colors.fig}
\end{figure}

Using photometry from the DR3 quasar catalog and the {\it GALEX}
object catalogs, we constructed the color-color diagrams in Figures
\ref{dr3_colors.fig} and \ref{uvopt_colors.fig}.  The top panel of
Figure \ref{dr3_colors.fig} shows the locations of SDSS stars (red
points and contours) and the quasar sample (black points and contours)
in $g-r$ vs.\ $u-g$ color-color space.  The bottom panel of Figure
\ref{dr3_colors.fig} shows only the 3004 quasars with detections in both
UV bands that comprise the clean sample, along with the same set of
SDSS stars.  Nearly all of the optically red quasars with $(u-g) > 0.7$
are not detected in both UV bands.  Note that the colors of the DR3
quasars with UV detections are uniformly distributed within the same
region that the quasar locus in the top panel occupies, indicating no
strong optical bias in our detections.

Using the same color scheme as Figure \ref{dr3_colors.fig}, Figure
\ref{uvopt_colors.fig} shows the UV/optical colors of the quasars
detected in both UV bands, along with {\it GALEX} stars from selected DIS
fields, which we matched to the SDSS database.  Since nearly all
of the DIS fields covering SDSS quasars are centered on the high
galactic latitude sky, the density of stars on the sky is relatively
low (especially in the FUV) and most of the FUV detections are near
the flux limit.  Since far fewer quasars have FUV detections and the
scatter in the UV photometry is large, the regions of ($n-u$) vs. ($f-n$)
color-color space occupied by stars and quasars are not as well defined.
For the bottom panel of Figure \ref{uvopt_colors.fig}, the NUV quasar
detections were used to construct the $u-g$ vs. $n-u$ colors.

The density of stars in the NUV for a given location on the sky is
higher than the corresponding density in the FUV by about an order of
magnitude, and thus we were able to match more {\it GALEX} stars to the
SDSS database.  The star-quasar separation in $u-g$ vs. $n-u$ color-color
space indicates that the inclusion of UV colors can aid in the color
selection of quasars using UV/optical photometry.  In addition, the loci
of points are consistent with reddening along the line of sight through
the intergalactic medium --- that is to say the bulk of the matches do
not have ``blue'' colors in the UV while having ``red'' colors for the
optical bands.

\subsection{Color-Redshift Relations} 
\label{colorzed}

\begin{figure}
\includegraphics[width=1.0\textwidth]{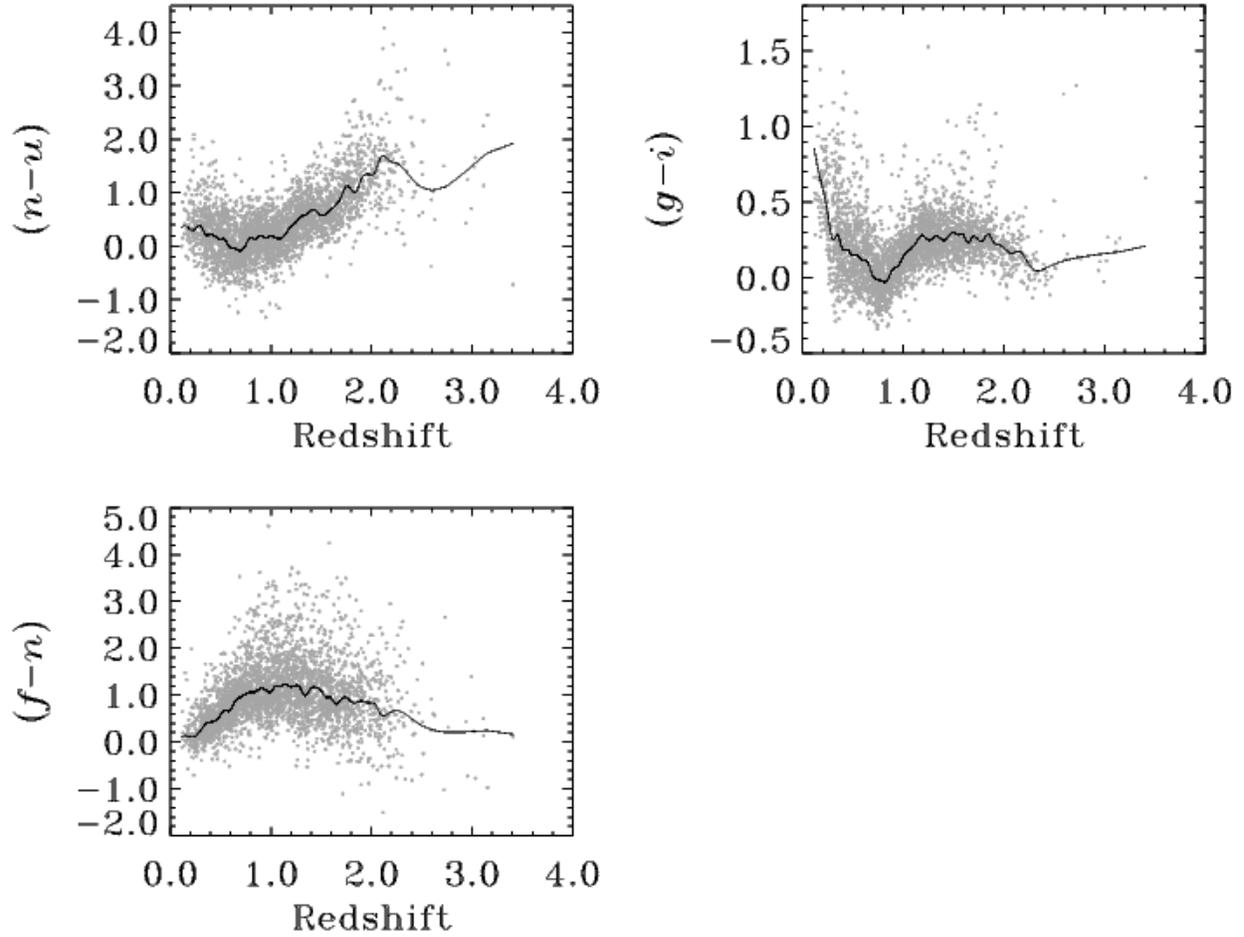}
\caption{UV/optical colors as a function of redshift for all of the UV
 detections.  The median color in each redshift bin was interpolated
 for each quasar at each redshift (black curves).  Galactic extinction
 corrections have been applied to all objects.}
\label{color_zedQ.fig}
\end{figure}

The observed structure of the quasar colors with redshift is quite complex
--- in addition to intergalactic reddening of the continuum, there are
strong emission lines (e.g., Balmer recombination lines, Mg II, CIV, and
Ly $\alpha$), and there is the SBB ($\sim 2000-4000$ {\AA}) due mostly to
Balmer continuum and Fe II complex emission.  Figure \ref{color_zedQ.fig}
shows the color-redshift relations for the UV/optical colors $f-n$, $n-u$,
and $g-i$ colors (all magnitudes corrected for Galactic extinction).
After binning the quasars by redshift, we have used a cubic spline
fit to interpolate and subtract the median color at each redshift to
obtain the relative colors.  Broad emission and/or absorption lines will
affect the quasar colors as they pass through the photometric bands,
and are primarily responsible for the observed color-redshift relations.
For example, in the top-left panel of Figure \ref{color_zedQ.fig} the
Ly $\alpha$ emission line has entered the NUV band by $z \sim 0.4$ and
exits at $z \sim 1.3$, after which the Ly $\alpha$ forest continues to
redden the UV/optical $n - u$ color with increasing redshift. Quasars that
display broad absorption lines in their spectra (BALQSOs) can have colors
that are either redder or bluer, depending on the location of absorption
troughs relative to the photometric filter bandpasses, and \citet{Reichard
2003} showed that the overall flux deficit due to reddening dominates
over the extinction due to the absorption troughs themselves.

\begin{figure}
\includegraphics[width=1.0\textwidth]{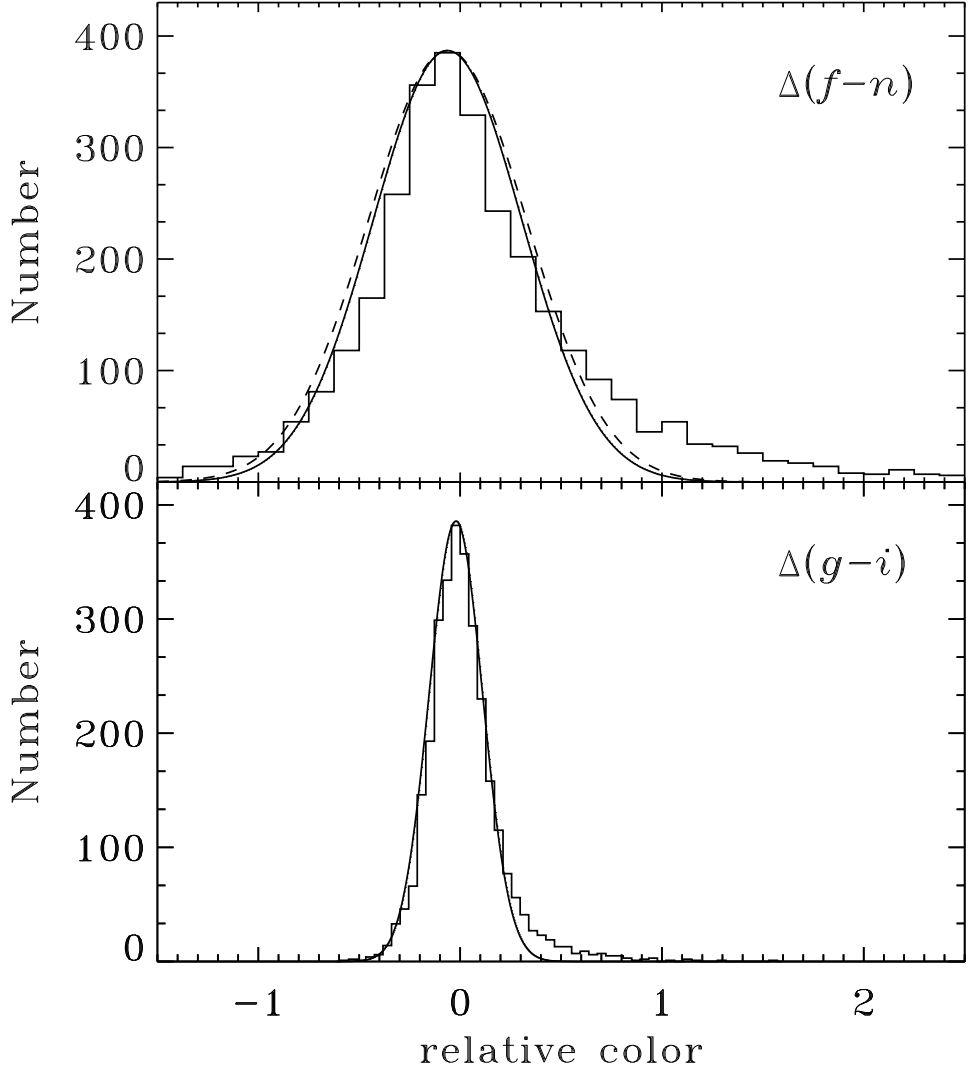}
\caption{The UV/optical relative colors for all of the UV detections,
 computed by subtracting the median color at each redshift.  The solid
 curves are the least-squares best-fit Gaussians to the peak and blue
 tail of each histogram.  Note the red tails of both distributions that
 rise above the fits (see \S\ref{colorzed}).  The red tail in $\Delta
 (g-i)$ is not produced by the SDSS photometric uncertainties, but
 is rather due to reddening by dust along our line of sight or at the
 redshift of the quasar \citep{Richards 2003, Hopkins 2004}.  The dashed
 curve in the top panel is the result of Monte Carlo simulations of the
 $\Delta (f-n)$ distribution using our empirical UV photometric errors
 (\S\ref{photometry}).  The comparable widths of both Gaussians indicates
 that the broader distribution of $f - n$ colors is not due to intrinsic
 differences in the UV SEDs of the detections, but is dominated by the
 photometic uncertainties.  However, the `excess' population of objects
 with large $\Delta (f-n)$, may suggest the presence of a subset of
 heavily reddened objects.}
\label{rel_color_hist.fig} 
\end{figure}

We select the distributions of the relative $f-n$ and relative $g-i$
colors (hereafter $\Delta$($f-n$) and $\Delta$($g-i$), respectively)
shown in Figure \ref{rel_color_hist.fig} for further analysis.  The solid
curves are the least-$\chi^2$ Gaussian fits to the blue side and peak of
each histogram.  \citet{Richards 2003} and \citet{Hopkins 2004} have shown
that the red tail on the $\Delta$($g-i$) distribution is not produced
by errors in the SDSS photometry, but is rather a true indicator of
reddening by dust along our line of sight at the redshift of the quasar.
The reddest quasars may not be detected in the NUV or FUV, and therefore
would not contribute to the observed relative color distribution in
Figure \ref{rel_color_hist.fig}.  However, the $\Delta(g-i)$ KS-test
(\S\ref{analysis}) seems to show that there is no significant bias with
respect to $\Delta$($g-i$).

The dashed curve on the top panel of Figure \ref{rel_color_hist.fig} is
the result from Monte Carlo simulations ($10^5$ runs) of $\Delta$
($f-n$) using a Gaussian distribution of photometric errors that use
the empirical UV magnitude-rms relations as determined in
\S\ref{photometry}.  Each best-fit Gaussian has $\sigma \sim 0.4$,
which is more than a factor of 2 larger than if the errors quoted in
the GR1 tables are used.  The blue and red curves in the top panel of
Figure \ref{rel_color_hist.fig} indicate the dispersion in $\Delta$
($f-n$) is dominated by photometric uncertainties, and that it is not
purely the result of intrinsic differences in the UV quasar SEDs.
However, the red tail of the $\Delta$($f-n$) distribution ($\sigma \sim
0.13$) suggests a marginally detected population of quasars heavily
reddened in the UV, which is also supported by a red tail observed in
the $\Delta$($n-u$) distribution.

\begin{figure}
\includegraphics[width=1.0\textwidth]{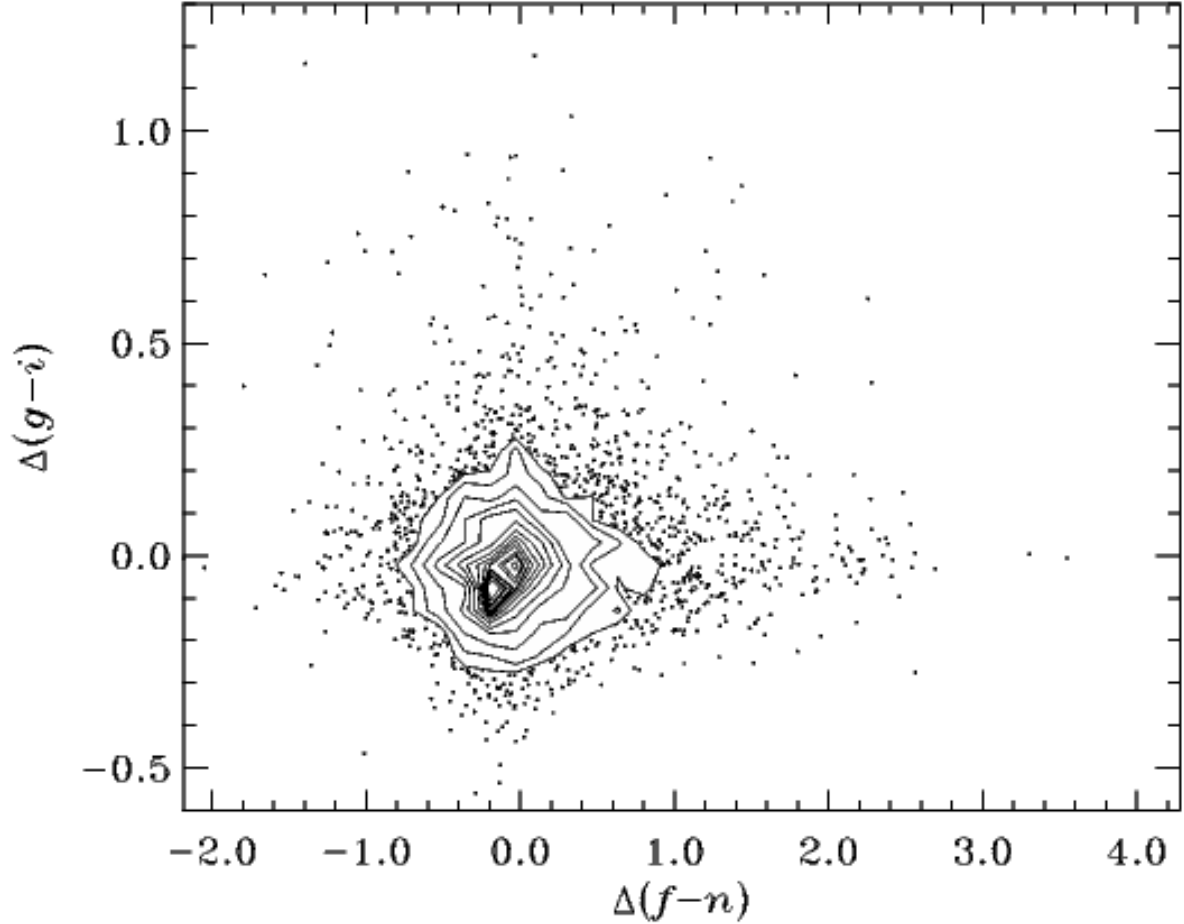}
\caption{The points and contours show the UV/optical relative colors
  for the quasar detections on both {\it GALEX} bands.  The Spearman
  Rank-Order Correlation results do not indicate any significant
  correlation (or anticorrelation) of the UV/optical relative colors
  (see \S\ref{colorzed}).}
\label{fngi_relcolor.fig}
\end{figure}

In terms of $g$-magnitude, $A_u$, and redshift, the quasars with both
red $\Delta$($f-n$) and $\Delta$($g-i$) colors are statistically
indistinguishable from those with bluer relative colors (see Figure
\ref{fngi_relcolor.fig}).  From computing the Spearman Rank-Order
correlation coefficients, the significance level ($\sim 0.88$)
indicates that it is unlikely the relative colors are correlated (or
anticorrelated) in any way.  The quasars with red UV colors in our
detections could be the result of intrinsically redder SEDs, the
Ly $\alpha$ forest, or due to the presence of emission or absorption
lines in one of the bands that can produce quasar colors that are
redder than the bulk of the quasars at that redshift.  Therefore,
while the use of UV colors may aid in the selection of dust-reddened
quasars, it will also select normal (but perhaps optically steep)
quasars \citep[see][and references therein]{Richards 2003}.

\subsection{SEDs}
\label{seds}

\begin{figure}
\includegraphics[width=1.0\textwidth]{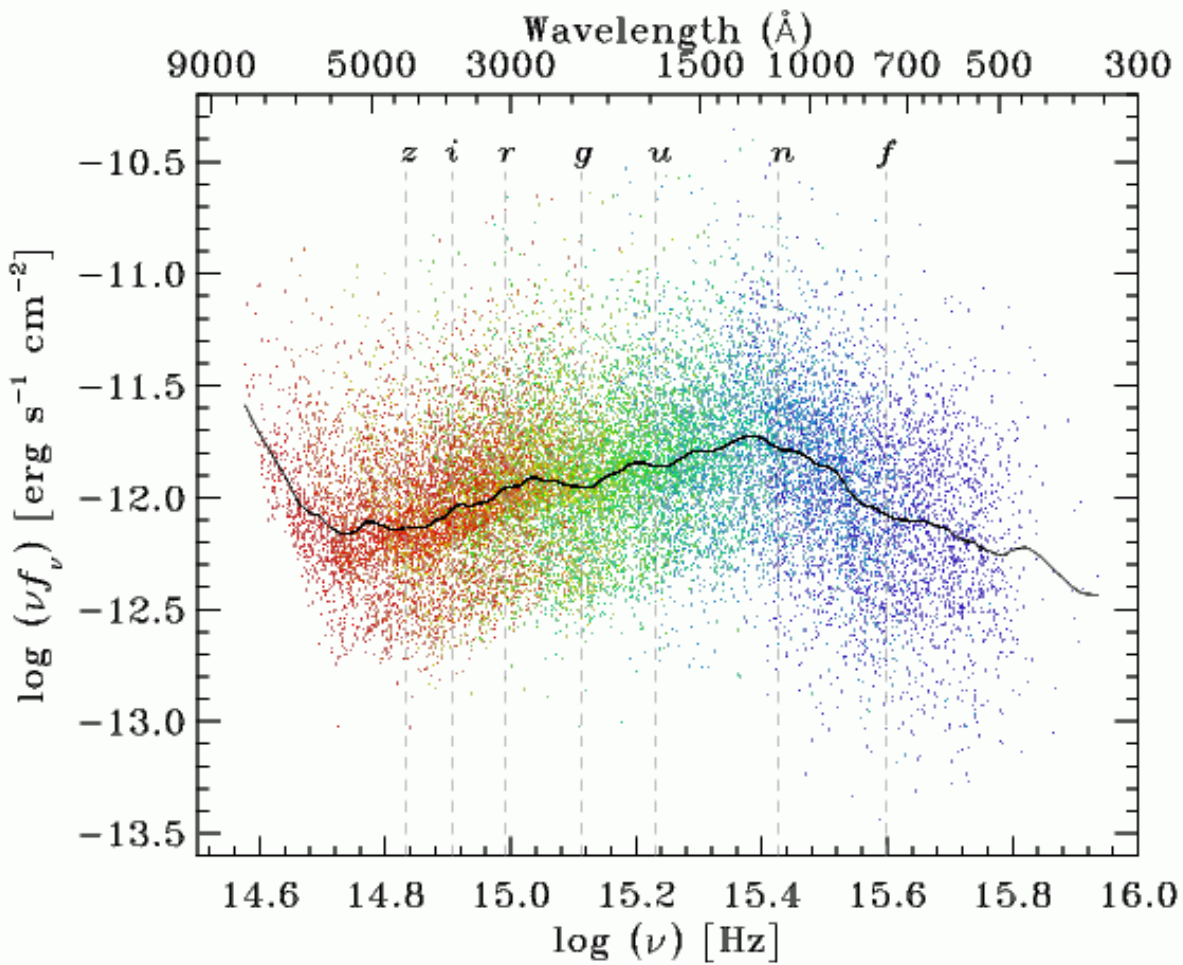}
\caption{Rest-frame spectral energy distributions (SEDs) for all of
 the detections (see \S{\ref{matching}}).  There is a one-to-one
 correspondence between the color of the plotting symbols and the
 effective wavelength of each band (e.g., the FUV is violet and the
 $z$-band is red).  The solid curve is the smoothed median SED using
 dlog($\nu$) = 0.01.  The dashed gray vertical lines correspond to the
 indicated photometric bands, shown with the effective frequency of
 each de-redshifted using the median redshift ($z \sim 1.02$) for all
 of the objects.}
\label{median_sed.fig}
\end{figure}

\begin{figure}
\includegraphics[width=1.0\textwidth]{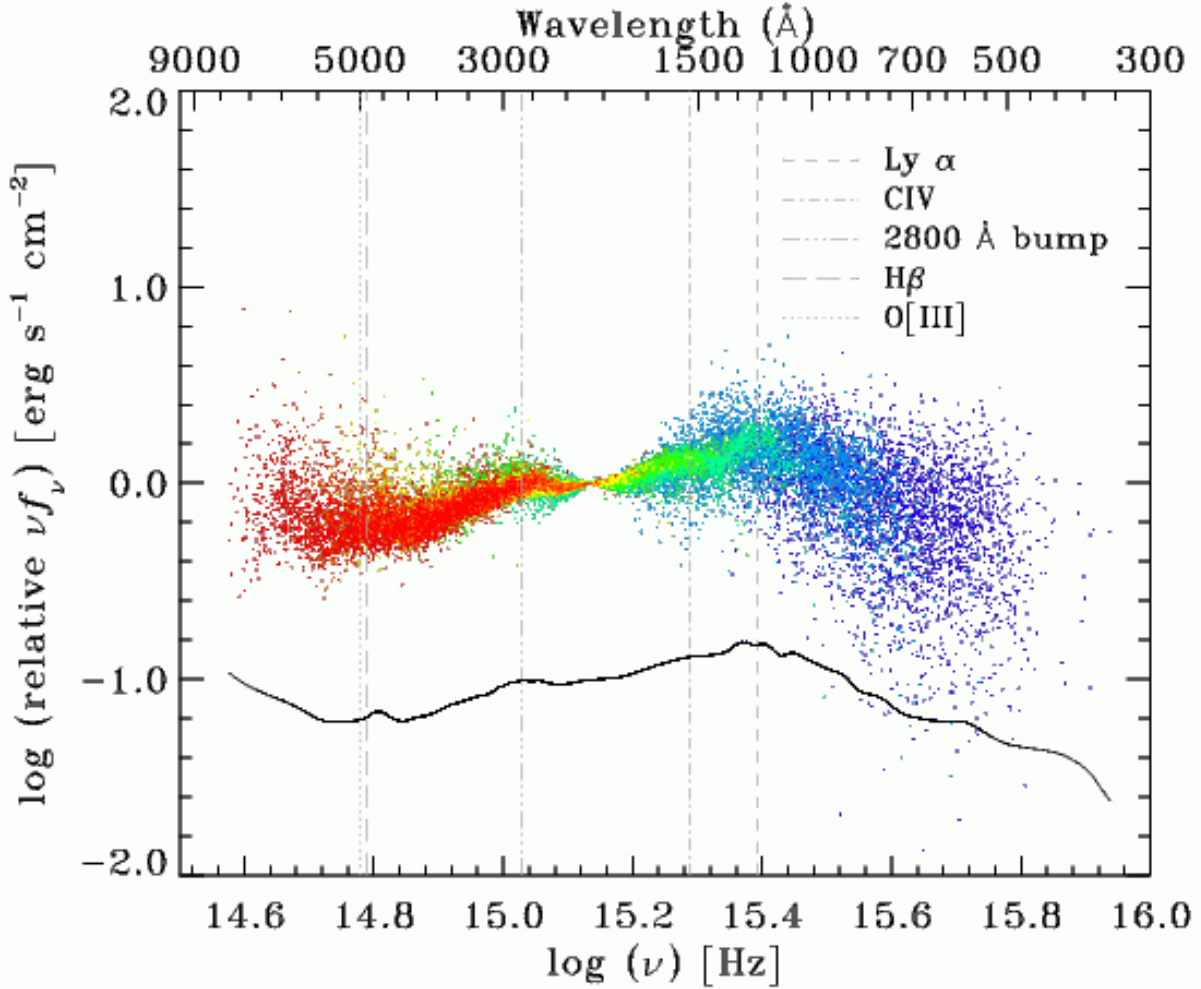}
\caption{Our rest-frame SEDs from Figure \ref{median_sed.fig} where
 each individual SED has been normalized to its own flux at 2200
 {\AA}.  The peak is near the location of Ly $\alpha$ (1216 {\AA}),
 where the overall rise in the relative flux is largely due to the
 presence of the Ly $\alpha$ emission line within the filter
 bandpasses.  The locations of other prominent features are also
 visible, such as C IV $1550$, the 2800 {\AA} bump, as well as
 H$\beta$ and O [III].}
\label{norm_sed.fig}
\end{figure}

After applying Galactic extinction corrections to the UV/optical
photometry and de-redshifting the effective frequency of each filter,
we constructed the rest-frame SEDs shown in Figure
\ref{median_sed.fig}.  The bluest points are the FUV photometry, with
the reddest points representing the reddest optical band (the SDSS-$z$
filter).  The black line is the median smoothed SED for all of the
detections, and the vertical dashed lines show the effective
frequencies of each photometric band, after de-redshifting using the
median redshift of all of the detections ($z = 1.01$).  Figure
\ref{norm_sed.fig} shows the result after each individual SED was
normalized to its own flux at 2200 {\AA} by linear interpolation in
log $(\nu f_{\nu})$ vs. log $(\nu)$ space.  A cubic spline
interpolated median SED is the curve shown below the colored points,
shifted downward 1 dex in logarithmic relative flux for clarity.  The
signatures of various canonical features known to be present in quasar
composite spectra (such as that of \citet{Vanden Berk 2001}) are
visible, denoted by reference lines given for the locations of Ly
$\alpha$, C IV $\lambda1550$, the 2800 {\AA} bump (Fe II complex, Mg
II, and Balmer continuum emission), H $\beta$, and O [III]
$\lambda5007$.

In addition to observational effects (e.g., line-of-sight absorption),
measurement uncertainties, and any intrinsic differences in the quasar
emission, variability also plays a role in the variety of SEDs present in
Figure \ref{norm_sed.fig}.  The anticorrelations between the variability
amplitude and wavelength, as well as luminosity \citep[see][and
references therein]{Vanden Berk 2004} render UV observations more
sensitive to changes in emergent flux levels.  Variability can therefore
not only displace quasars from their true locations in multidimensional
color-color space, but also increases the dispersion of the SEDs in
our sample.  In principle, foreground absorption systems can exclude
a significant fraction of the quasar population from flux-limited
surveys such as the SDSS due to attenuation in the rest-frame UV.
However, \citet{Akerman 2005} found no direct evidence for this effect
\citep[c.f.][]{York 2006}.  The scatter in Figure \ref{norm_sed.fig}
extending toward larger relative flux at longer wavelengths is primarily
due to the increasing contribution of the host-galaxy to the optical
flux for the lower redshift quasars, which dominate the SED near the
long wavelength limits of our composite SED.  The peak flux of the
SED occurs near $\sim 1000$ {\AA} or so, before the Ly $\alpha$ forest
begins to significantly extinguish the detected UV flux (Ly $\alpha$ has
been redshifted beyond the NUV band by $z \sim 1.33$).  The rest-frame
wavelength coverage of the UV/optical SEDs extends shortward to $\sim
350$ {\AA} for the highest redshift detection at $z = 3.41$ and longward
to $\sim 9000$ {\AA} for the lowest redshift detection at $z = 0.12$.
Table \ref{sample_sed.tab} illustrates the format and content of our
overall SED that accompanies the electronic edition of the journal.

\begin{figure}
\includegraphics[width=1.0\textwidth]{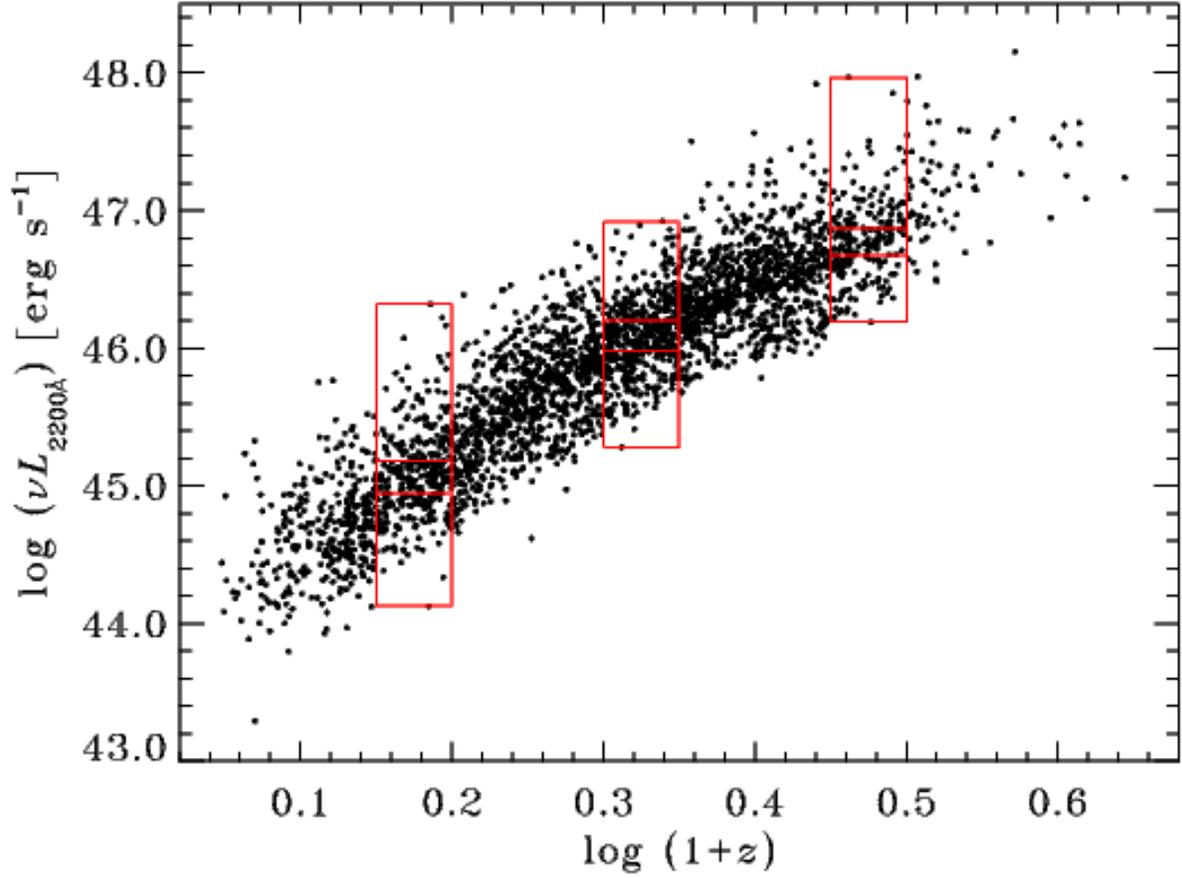}
\caption{The specific luminosity (at 2200 {\AA}; rest-frame)
 vs. redshift distribution of the 3004 NUV/FUV \textit{GALEX} detections
 of DR3 quasars.  The three highlighted redshift ranges correspond
 to 0.15 $<$ log ($1+z$) $<$ 0.20, 0.30 $<$ log ($1+z$) $<$ 0.35,
 and 0.45 $<$ log ($1+z$) $<$ 0.50, respectively.  We further divide
 the quasars within each redshift bin into three bins of luminosity,
 which are centered on the median value and with each luminosity bin
 containing a third of the quasars.  We have used the quasars in each
 bin to construct the UV/optical SEDs of Figure \ref{seds.fig}.}
\label{lrdistr.fig}
\end{figure}

\begin{figure}
\includegraphics[width=1.0\textwidth]{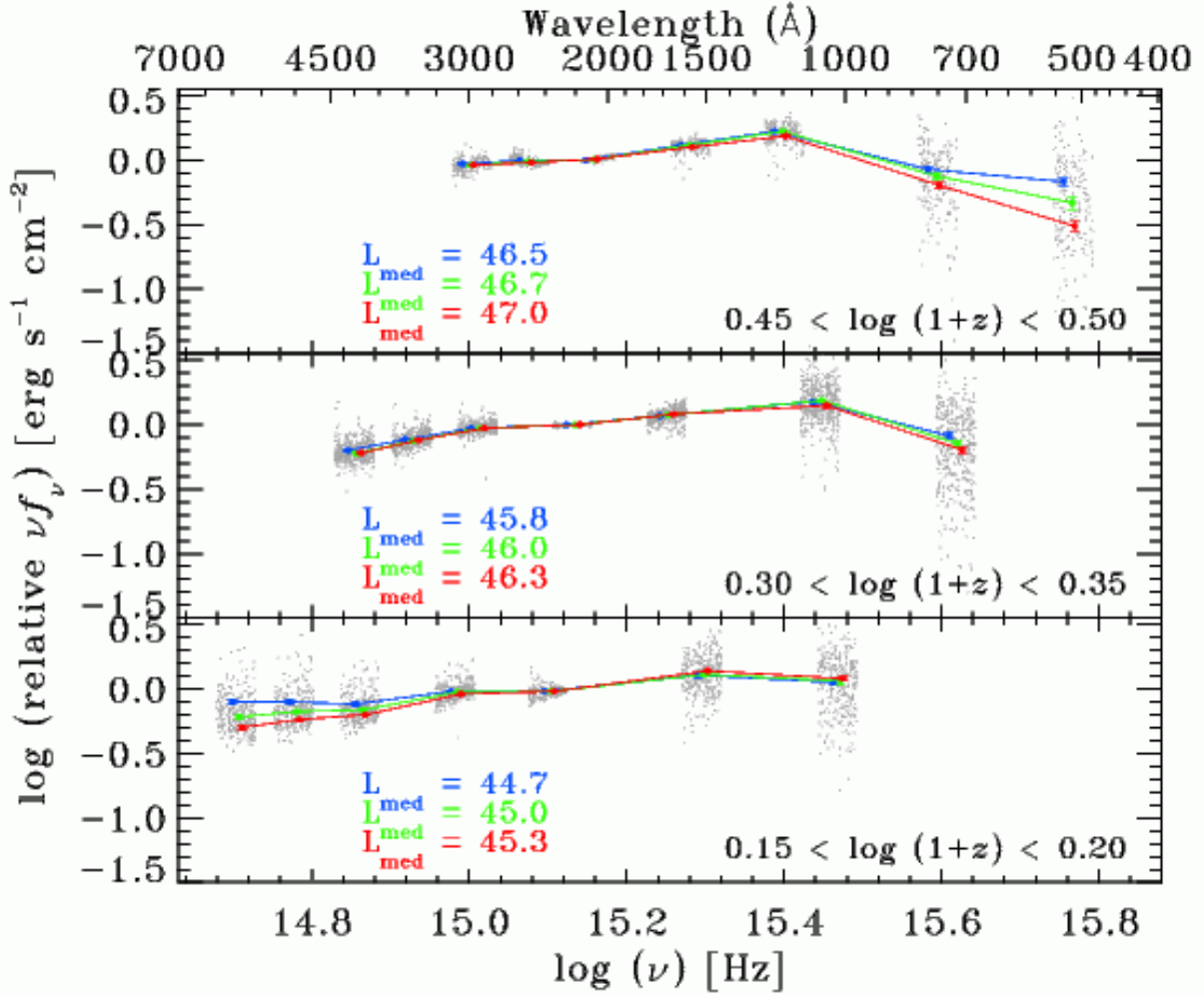}
\caption{SEDs in bins of redshift and luminosity.  The color points 
 and lines are the median normalized SEDs (see Figure \ref{norm_sed.fig})
 constructed from independent redshift and luminosity bins, drawn
 from the luminosity-redshift distribution of Figure \ref{lrdistr.fig}
 over the indicated redshift range.  The median log $(\nu$L$_{\nu})$
 at 2200 {\AA} of the quasars used to construct each individual SED are
 given by $L_{med}$.  In the top and middle panels, Lyman absorption
 and extragalactic extinction are the primary reasons for the SED
 turnover near $\log \nu \sim 15.4$ (Ly $\alpha$).  The error bars
 (comparable in size to the plotting symbols) are the 1$\sigma$ rms (the
 standard deviation normalized by $\sqrt{N}$, where $N$ is the number
 of objects in each bin) computed for all of the quasars (gray points)
 contributing to each respective SED.  The host-galaxy contamination
 becomes increasingly important at longer wavelengths, especially at lower
 redshifts where the correlation is easily visible in the bottom panel.
 In the top panel, our SEDs cover the shortest rest-frame wavelengths,
 where we see a strong anti-correlation of the overall strength of the
 UV continuum with luminosity.}
\label{seds.fig} \end{figure}

In order to constrain our understanding of the UV spectral slope, as well
as its redshift and luminosity dependence, one must take into account
various observational effects.  A significant fraction of the quasar
luminosity comes from broad emission lines, and Lyman absorption along our
line of sight can significantly alter the quasar colors (especially beyond
\textit{z} $\sim$ 1).  In addition, there is an artificial correlation
between quasar redshift and luminosity (e.g., Figure \ref{lrdistr.fig}),
and Lyman absorption strongly attenuates the SEDs.  The analysis of any
luminosity-dependence of the SEDs requires the sample to have a large
range in luminosity (and therefore redshift), and small redshift windows
allow for study of the luminosity-dependence as a function of wavelength.
In order to remove the effects of the emission lines, Lyman absorption
and to account for the $L-z$ correlation, we divide our UV detections
with redshifts $0.11 < z < 3.41$ into three redshift bins: (1) $0.15
<$ log $(1+z)$ $< 0.20,$ (2) $0.30 <$ log $(1+z)$ $< 0.35,$ and (3)
$0.45 <$ log $(1+z)$ $< 0.50$ (see the highlighted regions of the $L-z$
distribution in Figure \ref{lrdistr.fig}).  We then computed median
SEDs by binning the quasars in each redshift bin by their 2200 {\AA}
luminosity.  The resulting median UV/optical SEDs are shown in Figure
\ref{seds.fig}.  The three panels each correspond to the redshift range
indicated; the blue, red, and green points and lines represent different
luminosity bins --- the median monochromatic luminosity at 2200 {\AA}
for the quasars in each bin (denoted by $L_{med}$) is given in units
of log ($\nu L_{\mbox{\footnotesize 2200{\AA}}}$) and the error bars on
each median SED are the 1$\sigma$ rms (the standard deviation normalized
by $\sqrt{N}$, where $N$ is the number of objects) calculated for all
objects in each bin (gray points).  For several points, the errors in
the median are smaller than the plotting symbols.

The weak anticorrelation between the UV power-law spectral slope
($\alpha_{\nu}$) and the quasar luminosity for the lowest redshift bin
(bottom panel of Figure \ref{seds.fig}) is primarily due to
contamination by the host-galaxy, as the lower luminosity quasars are
preferentially at lower redshifts where the host-galaxy contribution
can be comparable to the nuclear emission.  A discussion of the
relationship between the AGN and host-galaxy fluxes of SDSS broad-line
active galaxies is given by \citet{Vanden Berk 2006}.  Our results
also reveal a significant anticorrelation between $\alpha_{\nu}$ and
luminosity for our highest redshift bin (top panel of Figure
\ref{seds.fig}) covering the shortest rest-frame wavelengths, in the
sense that the higher luminosity quasars have a steeper UV SED.  In
terms of the relative FUV flux for each median SED in the highest
redshift bin, the lowest luminosity ($L_{med} = 46.5$) UV continuum
level is 37\% higher (5.2$\sigma$) than for the most luminous
($L_{med} = 47.0$) median SED.

The observed luminosity-dependence of the UV SED shape at the shortest
wavelengths is consistent with idea that the Baldwin Effect
\citep[hereafter BE;][]{Baldwin 1977} may be due to intrinsic
differences in the strength of the ionizing continuum \citep[see][for
a review]{Osmer 1999}.  Other studies have also drawn similar
conclusions, while also taking into account the importance of
metallicity in determining the emission line strengths as functions of
the SED shape \citep[e.g.,][]{Korista 1998}.  For the higher
luminosity quasars (having apparently redder SEDs), fewer incident UV
photons are available to photoionize the broad-line gas and will
result in lower emission line equivalent widths.  Except for the
strong luminosity-dependence of our SEDs at the shortest wavelengths,
and the host-galaxy contributions seen at the longest wavelengths, the
median SEDs within each redshift bin are largely indistinguishable.

\section{DISCUSSION AND CONCLUSIONS} 
\label{summary}

We have presented the results of {\it GALEX} GR1 UV matches to
optically-selected quasars in the SDSS DR3 quasar catalog.  In the final
clean sample, 81.1\% of the DR3 quasars have an NUV detection, while
47.2\% have a detection in both the NUV and FUV bands.  An analysis of
the subsamples of the detections and non-detections indicates that the
non-detections tend to be optically faint objects or at high redshift.
The accuracy of the SDSS astrometry has allowed us to obtain a rigorous
estimate of the accuracy of {\it GALEX} positions --- showing them to
be accurate to 0.6-0.7{\arcsec}.   We have also carried out an empirical
analysis of the {\it GALEX} UV photometry using multi-epoch tiles from the
Deep Imaging and All-Sky Imaging Surveys that shows the true photometric
errors are a factor of $\sim2$ larger than those given in the GR1 object
tables.  The distributions of UV/optical colors of the quasar matches
indicate that the inclusion of UV/optical colors can aid in the color
selection of quasars, but also that large uncertainties in the {\it GALEX}
UV photometry prevents the use of UV colors alone to separate stars
from quasars.  In addition to other observational effects (e.g., Lyman
absorption), extinction within our own Galaxy is likely causing some of
the optically-selected quasars to be missed by {\it GALEX}, especially in
the FUV band.  Although the observed width of the relative UV $f-n$ color
distribution is dominated by photometric error, the red tail suggests
a marginally detected population of reddened quasars --- either due to
intervening absorption systems or due to dust at the quasar redshift.

In order to investigate a possible luminosity-dependence of the shape
of the UV quasar continuum, we have constructed rest-frame SEDs for 3004
quasars --- for which we can also make comparisons to theoretical models.
For example, \citet{Hubeny 2000} have used non-LTE accretion disk models
to predict the emergent SEDs, and their results predict both a change
in the slope of the power-law continuum in the vicinity of Ly $\alpha$
and a steepening of the continuum with increasing quasar luminosity (for
fixed values of $L/L_{\rm Edd}$).  To properly account for the correlation
between redshift and luminosity, we have analyzed our SEDs in independent
redshift and luminosity bins.  Our results show an anticorrelation between
$\alpha_{\nu}$ and the quasar luminosity at the shortest wavelengths
at the 5.2$\sigma$ level, which appears to be consistent with the idea
that the BE may be due to intrinsic differences in the strength of
the ionizing continuum.  The \citet{Scott 2004} results show a harder
UV continuum in their {\it FUSE} composite relative to the {\it HST}
composites of \citet{Zheng 1997} and \citet{Telfer 2002}, which display a
break in the spectral index near $\sim 1000$ {\AA}.  \citet{Scott 2004}
note that their sample of AGN has a median luminosity that is nearly
an order of magnitude lower than the sample of \citet{Telfer 2002}.
The anticorrelation we find in the UV between $\alpha_{\nu}$ and
luminosity is qualitatively consistent with the idea that the {\it FUSE}
composite having a significantly harder spectral index ($\alpha = 0.56$)
than the {\it HST} composite ($\alpha = 1.76$), is due to the lower
median luminosity of the {\it FUSE} sample.

Despite the range of $L/L_{\rm Edd}$ spanned by our sample, our
findings are also in good qualitative agreement with the model
predictions of \citet{Hubeny 2000}.  Our observed luminosity-dependent
UV SED shape is also consistent with the results of \citet{Shang 2005}
making use of both {\it FUSE} and {\it HST} spectra, who report a
softer FUV spectral index for the higher luminosity AGNs in their
sample.  However, \citet{Shang 2005} also note that most of their
objects display a spectral break near $\sim$ 1100 {\AA}, which is in
rough agreement with previous work using {\it HST} composite spectra
\citep[e.g.,][]{Telfer 2002}.  Therefore, although the \citet{Shang
2005} results seem to confirm the change in the continuum slope near
Ly $\alpha$ predicted by \citet{Hubeny 2000}, it remains unclear as to
whether or not the location of any UV spectral break is a function of
luminosity.

Some work has suggested that the shape of the ionizing continuum
may be correlated with metallicity, or that it may be the result
of luminosity-dependent spectral variations \citep[e.g.][]{Dietrich
2003}.  Differences in the Lyman absorption along different sightlines
are unlikely to cause the luminosity-dependence we see, since the
composite SEDs are averaged over a large number of objects at nearly
equal redshifts.  Our study has not taken into account the presence of
intervening absorption systems \citep[e.g.,][]{York 2006}.  The {\it
GALEX} photometric bands are wide, covering a significant range of
Lyman absorption --- a rigorous treatment of corrections that take
into account absorption by the Ly $\alpha$ forest, Lyman limit systems
(LLSs), and damped Ly $\alpha$ absorbers (DLAs) is beyond the scope of
this paper.  However, our large sample size has permitted detailed study
of the UV/optical regime without being highly sensitive to intervening
galaxy absorbers or lines of sight that intersect LLSs and/or DLAs.
Future SDSS and {\it GALEX} data releases will permit the construction of
even larger UV/optical samples of quasars.  In addition, the larger data
volume and the anticipated improvements in the photometric calibration
of {\it GALEX} data released in GR2 should increase the number of UV
detections of SDSS quasars, as well as aid the future studies that make
use of UV/optical colors in the color selection of quasars.

\acknowledgements 

The {\it GALEX} early release data was made available in 2003
December. The first major data release occurred on 2004 October 1,
coincident with the start of the {\it GALEX} guest investigator
program.  All {\it GALEX} data are served by the Multimission Archive
at the Space Telescope Science Institute at http://galex.stsci.edu/.
We gratefully acknowledge support from the National Aeronautics and
Space Administration (NASA) for the construction, operation, and
science analysis for the {\it GALEX} mission, developed in cooperation
with the Centre National d'Etudes Spatiales of France and the Korean
Ministry of Science and Technology. The grism, imaging window, and
uncoated aspheric corrector were supplied by France.  We acknowledge
the dedicated team of engineers, technicians, and administrative staff
from the Jet Propulsion Laboratory/Caltech, the Orbital Sciences
Corporation, the University of California at Berkeley, the Laboratory
Astrophysique Marseille, and the other institutions who made the
mission possible.

Funding for the SDSS and the SDSS-II has been provided by the Alfred
P. Sloan Foundation, the Participating Institutions, the National Science
Foundation (NSF), the United States Department of Energy, NASA, the
Japanese Monbukagakusho, the Max Planck Society, and the Higher Education
Funding Council for England.  The SDSS website is http://www.sdss.org/.
We also gratefully acknowledge support from NASA {\it GALEX} grants
428-09 58JA, NNGO5GE12G, and NNGO6GD03G, as well as NSF grant AST03-07582.

\newpage
\clearpage

\clearpage

\clearpage
\newpage
\begin{deluxetable}{cccc}
\centering 
\tabletypesize{\small} 
\tablewidth{0pt}
\tablecolumns{4}
\tablecaption{Best-Fit Parameters for Empirically Determined Photometric Errors}
\tablehead{
\colhead{Survey} & 
\colhead{Band} & 
\colhead{$a$} &
\colhead{$b$}
}
\startdata
AIS & NUV & 0.20525 & $-$4.7302 \\
AIS & FUV & 0.13360 & $-$3.0578 \\
DIS & NUV & 0.23466 & $-$5.7196 \\
DIS & FUV & 0.24413 & $-$5.8804 \\
\hline
\enddata
\tablecomments{
Parameters for the least-$\chi^2$ fits to the binned NUV and FUV rms
as a function of NUV and FUV magnitude using Equation \ref{emp_error}
in Figure \ref{photrms.fig}.
}
\label{phot_err.tab}
\end{deluxetable}

\clearpage
\newpage
\begin{deluxetable}{lccrcc}
\centering 
\tabletypesize{\small} 
\tablewidth{0pt}
\tablecolumns{6}
\tablecaption{Table for {\it GALEX} GR1 Matches to SDSS DR3 Quasars}
\tablehead{
\colhead{Object (SDSS J)} & 
\colhead{GALEX Object I.D.} & 
\colhead{R.A. (SDSS)} &
\colhead{Decl. (SDSS)} &
\colhead{Coord. Offset ({\arcsec})} &
\colhead{Redshift}
}
\startdata
000009.42$-$102751.9 & 2435007175665396402 & 0.039269 & $-$10.464428 & 1.62 & 1.8442 \\
000011.96+000225.3 & 2667127184236220309 & 0.049842 & 0.040372 & 0.44 & 0.4790 \\
000017.38$-$085123.7 & 2434901622551226166 & 0.072421 & $-$8.856607 & 1.17 & 1.2499 \\
000026.29+134604.6 & 2665121679322122922 & 0.109578 & 13.767970 & 0.65 & 0.7678 \\
000028.82$-$102755.7 & 2435007175666450605 & 0.120086 & $-$10.465496 & 0.32 & 1.1377 \\
\hline
\enddata
\tablecomments{
A full version of the above table accompanies the electronic edition
of the journal.  Table \ref{sample_table.tab} serves as a reference
for its final form and content.  All object coordinates are given in
decimal degrees (J2000.0).
}
\label{sample_table.tab}
\end{deluxetable}

\clearpage
\newpage
\begin{deluxetable}{ccl}
\centering 
\tabletypesize{\small} 
\tablewidth{0pt}
\tablecolumns{3}
\tablecaption{Table Format for \textit{GALEX} GR1 Matches to SDSS DR3 Quasars}
\tablehead{
\colhead{Column} & 
\colhead{Data Format} & 
\colhead{Description}}
\startdata
1 & A18 & SDSS DR3 Quasar Catalog object designation in hhmmss.ss+ddmmss.s (J2000.0) \\
2 & A19 & GALEX GR1 merged object catalog designation; -999 indicates no detection \\
3 & F11.6 & SDSS right ascension in degress (J2000.0) \\
4 & F11.6 & SDSS declination in degrees (J2000.0) \\
5 & F11.6 & GALEX right ascension (degrees) \\
6 & F11.6 & GALEX declination (degrees) \\
7 & F7.2 & Coordinate offset between GALEX/SDSS positions given in arcseconds \\ 
8 & F7.4 & redshift (SDSS) \\
9 & F7.3 & SDSS $u$ magnitude (not corrected for Galactic extinction) \\
10 & F6.3 & error in SDSS PSF $u$ magnitude \\
11 & F7.3 & SDSS $g$ magnitude (not corrected for Galactic extinction) \\
12 & F6.3 & error in SDSS PSF $g$ magnitude \\
13 & F7.3 & SDSS $r$ magnitude (not corrected for Galactic extinction) \\
14 & F6.3 & error in SDSS PSF $r$ magnitude \\
15 & F7.3 & SDSS $i$ magnitude (not corrected for Galactic extinction) \\
16 & F6.3 & error in SDSS PSF $i$ magnitude \\
17 & F7.3 & SDSS $z$ magnitude (not corrected for Galactic extinction) \\
18 & F6.3 & error in SDSS PSF $z$ magnitude \\
19 & F9.3 & GALEX $f$ magnitude (not corrected for Galactic extinction); -999 if no detection \\
20 & F8.3 & error in PSF $f$ magnitude (from GR1 object catalog); -999 if no detection \\
21 & F8.3 & error in PSF $f$ magnitude (empirical calibration) \\
22 & F9.3 & GALEX $n$ magnitude (not corrected for Galactic extinction); -999 if no detection \\
23 & F8.3 & error in PSF $n$ magnitude (from GR1 object catalog); -999 if no detection \\
24 & F8.3 & error in PSF $n$ magnitude (empirical calibration) \\
25 & F7.3 & Galactic extinction in the $u$ band ($A_u$) \\
26 & F7.3 & Logarithmic Galactic H I column density \\
27 & F8.3 & $M_i$ ($H_0 = 70$ km s$^{-1}$ Mpc$^{-1}, \Omega_M = 0.3, \Omega_{\Lambda} = 0.7, \alpha_{\nu} = -0.5$)* \\
28 & F8.3 & L$_{\nu}$ at 2200{\AA} (rest-frame; -999 if no coverage) \\ 
29 & F8.3 & L$_{\nu}$ at 5100{\AA} (rest-frame; -999 if no coverage) \\ 
30 & F6.3 & $\Delta(g-i)$ relative color (corrected for Galactic extinction) \\
31 & F8.3 & $\Delta(f-n)$ relative color (corrected for Galactic extinction) \\
32 & F7.5 & object distance from GALEX field center (degrees) \\
33 & F8.3 & GALEX near-UV effective exposure time (seconds) \\
34 & F8.3 & GALEX far-UV effective exposure time (seconds) \\
35 & I1 & GALEX survey identifier (4 = AIS, 3 = NGS, 2 = MIS, 1 = DIS) \\
\hline
\enddata
\tablecomments{
Data format and content for the full version of Table
\ref{sample_table.tab} that accompanies the electronic edition of the
journal. 
}
\label{table.tab}
\end{deluxetable}

\begin{deluxetable}{llccccc}
\tabletypesize{\small} 
\tablewidth{0pt}
\tablecolumns{7}
\tablecaption{Kolmogorov-Smirnov Test Results Summary}
\tablehead{
\colhead{Sample 1} &
\colhead{Sample 2} & 
\colhead{$A_{u}$} & 
\colhead{$g$-magnitude} & 
\colhead{$\Delta(g-i)$} &
\colhead{Redshift} &
\colhead{NUV $t_{\rm eff}$}
}
\startdata
NUV$+$FUV detections & DR3 Sample & 0.18 & $< 10^{-12}$ & 0.69 & $< 10^{-12}$ & $< 10^{-12}$ \\
NUV detections & DR3 Sample & 0.80 & $4.3 \times 10^{-12}$ & 0.67 & $< 10^{-12}$ & $< 10^{-12}$ \\
no UV detection & DR3 Sample & $3.8 \times 10^{-3}$ & $< 10^{-12}$ & 0.63 & $< 10^{-12}$ & $< 10^{-12}$ \\
NUV$+$FUV detections & no UV detection &  $2.5 \times 10^{-3}$ & $< 10^{-12}$ & 0.27 & $< 10^{-12}$ & $< 10^{-12}$ \\
NUV detections & no UV detection &  $3.4 \times 10^{-4}$ & $< 10^{-12}$ & 0.62 & $< 10^{-12}$ & $< 10^{-12}$ \\
\hline
\enddata
\tablecomments{
Our Kolmogorov-Smirnov significance parameters computed for the
indicated subsample distributions of selected parameters, after
matches with $\theta > $ 2\farcs6 have been removed.
}
\label{ks_table.tab}
\end{deluxetable}

\clearpage
\newpage
\begin{deluxetable}{cccc}
\centering 
\tabletypesize{\small} 
\tablewidth{0pt}
\tablecolumns{4}
\tablecaption{Sample Table for Median SED}
\tablehead{
\colhead{Log $\lambda$ ({\AA})} & 
\colhead{Log $\nu$ (Hz)} & 
\colhead{$\nu f_{\nu}$} &
\colhead{rms}
}
\startdata
3.831 & 14.646 & 0.810 & 0.041 \\
3.788 & 14.689 & 0.699 & 0.030 \\
3.752 & 14.725 & 0.611 & 0.018 \\
3.731 & 14.747 & 0.610 & 0.026 \\
\hline
\enddata
\tablecomments{
A full version of the above table accompanies the electronic edition
of the journal.  Table \ref{sample_sed.tab} serves as a reference for
its final form and content. 
}
\label{sample_sed.tab}
\end{deluxetable}

\end{document}